\documentclass[12pt]{article}
\pdfoutput=1
\usepackage{graphicx} 
\usepackage{jheppub_X}
\usepackage{feynmp-auto}
\usepackage{color}
\usepackage{simpler-wick}
\usepackage{setspace}

\usepackage{cleveref}

\usepackage{tikz}
\usetikzlibrary{circuits.ee.IEC}

\newcommand\Ground{%
\mathbin{\text{\begin{tikzpicture}[circuit ee IEC,yscale=0.9,xscale=0.9]
\draw [line width=0.3mm ] (0,2ex) to (0,0) node[ground,rotate=-90,xshift=.65ex] {};
\end{tikzpicture}}}%
}

\crefname{table}{Table}{Tables}
\crefname{equation}{Eq.}{Eqs.}
\crefname{appendix}{App.}{Apps.}
\crefname{section}{Sec.}{Secs.}
\crefname{figure}{Fig.}{Figs.}

\makeatletter
\g@addto@macro\bfseries{\boldmath}
\makeatother

\newcommand{\s}{\hspace{0.8pt}}

\newcommand{\dd}{\text{d}}

\newcommand{\nn}{\nonumber \\}
\newcommand{\Lagr}{\mathcal{L}}

\definecolor{colorTC}{rgb}{.2,.7,.2}

\preprint{CERN-TH-2024-216 \newline \hspace*{\fill} YITP-SB-2024-35}

\title{Field Redefinitions Can Be Nonlocal}
\author[a,b,c]{Timothy Cohen,}
\author[d,e]{Matthew Forslund,}
\author[a]{and Andreas Helset\s}

\affiliation[a]{Theoretical Physics Department, CERN, 1211 Geneva, Switzerland}
\affiliation[b]{Theoretical Particle Physics Laboratory, EPFL, 1015 Lausanne, Switzerland}
\affiliation[c]{Institute for Fundamental Science, University of Oregon, Eugene, OR 97403, USA}
\affiliation[d]{C.\,N.\,Yang Institute for Theoretical Physics, Stony
Brook University, Stony Brook, NY, 11794, USA}
\affiliation[e]{
Physics Department,
Brookhaven National Laboratory, Upton, NY 11973, USA}

\emailAdd{tim.cohen@cern.ch}
\emailAdd{matthew.forslund@stonybrook.edu}
\emailAdd{andreas.helset@cern.ch}

\abstract{
We revisit the lore establishing the allowed space of field redefinitions and show that there are essentially no restrictions. 
Our conclusions hold to all orders in perturbation theory and for any dispersion relation.  
Field redefinitions can be nonlocal, symmetry breaking, or in certain cases have explicit dependence on spacetime. 
We address field redefinitions that can be resummed into the propagator, which demonstrates how
to perform perturbative calculations away from the minimum in field space.
Field redefinitions are used to derive higher-order Schwinger-Dyson equations, which imply multiparticle soft theorems.  
Non-standard field redefinitions are showcased using both relativistic and nonrelativistic examples.
}

\begin{document}
\maketitle
\flushbottom
\setcounter{page}{2}
%\newpage

\begin{spacing}{1.1}
\parskip=0ex

\section{Introduction}
Field redefinitions play a foundational role in the formulation of quantum field theory.  
The crucial statement is that $S$-matrix elements are invariant under field redefinitions, which has profound consequences for mapping from Lagrangians to physical predictions.  
They were first introduced in the context of gauge theories~\cite{Chisholm:1961tha, Kamefuchi:1961sb, tHooft:1972qbu, 'tHooft:962348,tHooft:1973wag}, since gauge transformations are themselves field redefinitions.  
They were then put to work again in the context of understanding the relation between the linear and nonlinear sigma model~\cite{Coleman:1969sm,Callan:1969sn}.  
They are a critical tool for understanding Effective Field Theories (EFTs), where field redefinitions underlie the trick to remove derivative interactions by applying the equations of motion~\cite{Deans:1978wn,Politzer:1980me,Arzt:1993gz,Passarino:2016saj,Manohar:2018aog,Criado:2018sdb}. 
In recent years, the connection between field redefinitions and coordinate changes on a field manifold~\cite{Honerkamp:1971sh, Volkov:1973vd, Tataru:1975ys, Alvarez-Gaume:1981exa, Alvarez-Gaume:1981exv, Vilkovisky:1984st, DeWitt:1984sjp, Gaillard:1985uh, DeWitt:1985sg} has been exploited to expose interesting physical consequences, both in the scalar sector of the Standard Model~\cite{Alonso:2015fsp, Alonso:2016btr, Alonso:2016oah, Nagai:2019tgi, Helset:2020yio, Cohen:2020xca, Cohen:2021ucp, Alonso:2021rac, Banta:2021dek, Talbert:2022unj, Alonso:2023jsi} and for scattering amplitudes of scalars and higher-spin particles~\cite{Finn:2019aip, Finn:2020nvn, Cheung:2021yog, Alonso:2022ffe, Helset:2022tlf, Helset:2022pde, Assi:2023zid, Jenkins:2023rtg, Jenkins:2023bls, Gattus:2023gep, Alonso:2023upf, Gattus:2024ird,Derda:2024jvo,Helset:2024vle}.

Given this wide range of applications, it is critically important to delineate the space of allowed field redefinitions.  
It is often assumed in the Standard Model EFT literature that in order for field redefinitions to leave the $S$-matrix invariant, they should include a linear term and be local functions of fields and derivatives.  
However, in the context of gauge transformations, 't Hooft and Veltman had long ago understood that nonlocal field redefinitions are completely allowed~\cite{tHooft:1973wag}.  
Additionally, mode-based EFTs such as Heavy Quark Effective Theory rely on time-dependent field redefinitions when separating the light from heavy modes \cite{Georgi:1990um}. 
Nonlocal field redefinitions are also common in the Soft Collinear Effective Theory (SCET) literature~ \cite{Bauer:2000yr,Bauer:2001yt,Beneke:2002ph,Beneke:2002ni}.
Our goal in this paper is to unify all of these different points of view into a coherent and complete picture.  
This will allow us to delineate in a precise way how to track the implications of field redefinitions, and to show how their effects cancel systematically in the language of Feynman diagrams. 
By understanding the details of the invariance of physical predictions under changes of field basis, we further the development of this abstract sentiment into a practical tool.

In this paper, we bring the diagrammatic proofs by 't Hooft and Veltman \cite{tHooft:1973wag} and Lam~\cite{Lam:1973qa} into the current zeitgeist and extend their argument to generic EFTs, including nonrelativistic theories.  
Furthermore, we show that in certain cases field redefinitions that are explicit functions of the spacetime coordinates are also allowed.  
This justifies the types of manipulation that are typically used when defining nonrelativistic EFTs.  
Symmetries may also be totally obscured by the change of field basis with no physical consequence. 
Our proof shows that one is free to perform field redefinitions with wild abandon, as long as care is taken to incorporate the full set of features that emerge in the redefined theory. 
This adheres to the EFT philosophy, where we need to consistently include all effects that appear at a given order in a power-counting parameter.

The rest of this paper is organized as follows. 
In \cref{sec:proof} we present a diagrammatic proof of field-redefinition invariance, first for a theory of a single real scalar field before generalizing the proof to a theory with arbitrary field content. 
From this we derive various implications for correlation functions and $S$-matrix elements, and obtain the Schwinger-Dyson equations in \cref{sec:Implications}. 
We then showcase the extended space of allowed field redefinitions in relativistic and nonrelativistic examples in \cref{sec:Examples}, before concluding in \cref{sec:Outlook}.

\section{The Diagrammatics of Field Redefinitions}\label{sec:proof}

We present a diagrammatic proof of field-redefinition invariance for $S$-matrix elements and correlation functions valid to all orders in perturbation theory.
For simplicity, the proof is given for a theory of a real scalar field, before discussing how the proof extends to general theories. 
The mechanism for the cancellation between terms is generic and holds for both relativistic and nonrelativistic theories with an arbitrary field content. 
The restrictions on what field redefinitions are acceptable are quite lenient: functions of spacetime derivatives and spacetime coordinates are allowed without any regard for locality or symmetry.

\subsection{Perturbative Field Redefinitions}\label{sec:perturbative}

We work with a general theory of a single scalar field 
\begin{equation}\label{eq:pathint}
    Z_\phi[J] = \int\! \mathcal{D}\phi\, \exp\!\bigg( i\s S[\phi] + i\! \int\! \dd^4x J(x) \phi(x) \bigg) \, ,
\end{equation}
where $S[\phi] = \int \dd^4 x \, \mathcal{L}[\phi].$
The Lagrangian is given by
\begin{equation}
    \mathcal{L}[\phi] = -\frac{1}{2} \phi\s \Delta^{-1}_x \phi + \mathcal{L}_\text{int}[\phi] \, ,
\end{equation}
where $\mathcal{L}_\text{int}[\phi]$ contains arbitrary interactions and spacetime derivatives of $\phi$, and the quadratic part of the Lagrangian determines the inverse propagator $\Delta_x^{-1}$. 
For example, a free massive scalar has $\Delta_x^{-1} = \square + m^2$. 
The arguments to follow will not depend on the form of $\Delta_x^{-1}$, so we will leave it general except in the case of specific examples. 

A general perturbative field redefinition takes the form
\begin{equation}
    \phi \rightarrow \widetilde{\phi}[\phi] = \phi + \lambda\s G[\phi] \,,
\end{equation}
where $\widetilde{\phi}[\phi]$ is assumed to be an invertible polynomial function of $\phi$, but we allow for an arbitrary dependence on spacetime derivatives (both in the numerator and denominator) and certain explicit functions of spacetime.
To simplify our analysis, we assume that the resulting vertices do not have any explicit dependence on spacetime, even though the field redefinition could have. This is the typical case for HQET as well as an example for nonrelativistic EFTs presented in Section~\ref{sec:Examples}.
% We assume that the Fourier transform of $G[\phi]$ exists.\footnote{Nonlocal operators in $x$ are equivalent to operators with infinitely many derivatives, see, e.g., Refs.~\cite{Tomboulis:2015gfa,Buoninfante:2018mre}. These cases are covered by our proof.}
We will track the effect of the field redefinition as a perturbative expansion in the parameter $\lambda$.

After performing this field redefinition, the generating functional becomes
\begin{align}
    Z_{\widetilde{\phi}}[J] &= \int \mathcal{D}\widetilde{\phi} \exp\!{\left( i\s S[\widetilde{\phi}] + i \int \dd^4x J(x) \widetilde{\phi}(x) \right)}\,, \notag\\[4pt]
    & = \int\mathcal{D}\phi \det\!{\bigg(\frac{\delta \widetilde{\phi} }{\delta \phi}\bigg) \exp\!{\left( i\s S\big[\phi+G\left[\phi\right]\big] + i \int \dd^4x J(x) \big[\phi(x)+G\left[\phi(x)\right] \big]\right)}}\,,
\label{eq:Zj}
\end{align}
where the new Lagrangian is
\begin{equation}\label{eq:lagrPrime}
    \mathcal{L}\big[\phi+G[\phi]\big] = -\frac{1}{2} \phi \Delta_x^{-1} \phi - \lambda G[\phi]  \Delta_x^{-1} \phi - \frac{\lambda^2}{2} G[\phi]  \Delta_x^{-1} G[\phi] + \mathcal{L}_\text{int}\big[\phi + \lambda G[\phi]\big] \, .
\end{equation}
The field redefinition has introduced a number of new terms in the generating functional.  
The goal of this section is to give a perturbative diagrammatic argument that 
\begin{align}
Z_{\vphantom{\tilde{\phi}}\phi}[J] = Z_{\widetilde{\phi}}[J]\,.
\label{eq:mainPoint}
\end{align}
We will then explore some implications of \cref{eq:mainPoint}.

In what follows, we will work in momentum space and assume that the Fourier transforms of $G$, $\Delta^{-1}$, and $\mathcal{L}_\text{int}$ exist. 
The only additional assumption required in this section is that the momentum conserving delta function in the Feynman rules is present. 
This is always the case when the vertices are independent of the spacetime coordinate, as we have assumed. 
It is worth noting that if $G$, $\Delta^{-1}$, and $\mathcal{L}_\text{int}$ are explicit functions of spacetime but may be rewritten in terms of infinitely many derivatives~\cite{Tomboulis:2015gfa,Buoninfante:2018mre}, these arguments still hold.
Extending the argument further may be possible working in position space, though this is beyond our scope.

The idea of this proof is to enumerate all possible Wick contractions of $\phi$ (denoted with an overbar as usual) that come from the term $(G\Delta^{-1}\phi)$. We will argue that every new diagram involving $(G\Delta^{-1}\phi)$ has a corresponding diagram that cancels it. 
Furthermore, the cancellation occurs fully off-shell. Therefore, any quantity built from these diagrams, such as correlation functions or $S$-matrix elements, will remain independent of $G(\phi)$ and thus invariant under field redefinitions. 

There are four different cases we must consider:
\begin{itemize}
    \item[(a)] The $(G\Delta^{-1} G)$ term in the Lagrangian in \cref{eq:lagrPrime} cancels with $\wick{(G\Delta^{-1} \c \phi) (G\Delta^{-1} \c \phi)}$.
    \item[(b)] The $n^\text{th}$ term in the Taylor series of $\mathcal{L}_\text{int}[\phi + \lambda G[\phi]] = \sum_{n} \frac{1}{n!}\mathcal{L}_\text{int}^{(n)} G^n$ cancels with contractions of the form $\wick{(G\Delta^{-1} \c\phi)^{n}\frac{1}{n!}\mathcal{L}_\text{int}^{(n)} \c\phi^n}$.
    \item[(c)] The term $(GJ)$ cancels with the contraction $\wick{(G\Delta^{-1} \c\phi) (J\c\phi)}$.
    \item[(d)] Loops of $(G\Delta^{-1} \phi)$ cancel with ghost loops from the determinant in \cref{eq:Zj}.
\end{itemize}
For all four cases, it should be understood that the vertices we consider may be a small part of a much larger diagram with arbitrary loops and external legs.
Since these cancellations happen fully off-shell for every vertex individually, this accounts for every new diagram to all loop orders proportional to any power $\lambda^{n}$. 

To first establish some notation that will be relevant for all four cases, we will use plain solid lines to indicate the field $\phi$, while clusters of red lines indicate the redefinition $G[\phi]$: 
\begin{equation}
\begin{fmffile}{phi}
\parbox{60pt}{
\begin{fmfgraph*}(40,40)
\fmfleft{i1}
\fmfright{o1}
\fmfv{decor.shape=circle,decor.filled=30,decor.size=20}{i1}
\fmf{plain}{i1,o1}
\fmfv{l={$\phi$}}{o1}
\end{fmfgraph*}}
\end{fmffile} , 
\qquad \text{and}\qquad \qquad 
\begin{fmffile}{Gphi}
\parbox{60pt}{
\begin{fmfgraph*}(40,40)
\fmfwizard
\fmfright{i1,i2,i3,i4,i5}
\fmfleft{o1}
\fmfblob{0.15w}{o1}
\fmf{plain,foreground=(.9,,0.1,,0.1),tension=1}{i2,o1}
\fmf{phantom,foreground=(.9,,0.1,,0.1),tension=1}{i3,o1}
\fmf{plain,foreground=(.9,,0.1,,0.1),tension=1}{i4,o1}
\fmf{dots,tension=0}{i2,i3}
\fmf{dots,tension=0}{i4,i3}
\fmfv{l={$G[\phi]$}}{i3}
\fmfv{decor.shape=circle,decor.filled=30,decor.size=20}{o1}
\end{fmfgraph*}} 
\end{fmffile} \hspace{12pt}.
\end{equation}
Here the blob is arbitrary for now.
We assume for simplicity that $G[\phi]$ only contains a single term with $n$ factors of $\phi$, though we write the equations to follow such that they may be generalized straightforwardly to accommodate versions of $G$ with multiple terms.
Arrows on $\phi$ legs indicate that the leg is acted on by the inverse propagator $\Delta^{-1}$:
\begin{equation}
\begin{fmffile}{phi_arrow}
\parbox{70pt}{
\begin{fmfgraph*}(50,40)
\fmfleft{i1}
\fmfright{o1}
\fmfv{decor.shape=circle,decor.filled=30,decor.size=20}{i1}
\fmf{plain,label.side=left}{i1,v1}
\fmf{phantom_arrow,label=$\Delta^{-1}$,label.side=left,tension=0,foreground=(.9,,0.2,,0.9)}{i1,v1}
\fmf{plain,tension=2}{v1,o1}
\fmfv{l={$\phi$}}{o1}
\end{fmfgraph*}}
\end{fmffile} .
\end{equation}
We will use green blobs $\ \ \begin{fmffile}{greenblob}
\parbox{9pt}{
\begin{fmfgraph*}(10,10)
\fmfleft{i1}
\fmfv{decor.shape=circle,decor.filled=30,decor.size=10,foreground=(.1,,0.5,,0.1)}{i1}
\end{fmfgraph*}}
\end{fmffile}$ for vertices coming from $\mathcal{L}_{\text{int}}$ and dashed blobs $\ \ \begin{fmffile}{dashedblob}
\parbox{9pt}{
\begin{fmfgraph*}(10,10)
\fmfleft{i1}
\fmfblob{10}{i1}
\end{fmfgraph*}}
\end{fmffile}$ for vertices coming from the $G$ and $G^2$ terms in \cref{eq:lagrPrime}.
We now explain each case in more detail.

\subsubsection*{Case (a)}
We will treat the field redefinition perturbatively, so that the propagator for $\phi$ is unchanged and any linear or constant pieces in $G[\phi]$ appear as insertions.
We therefore have
\begin{equation}
\begin{fmffile}{prop}
\parbox{60pt}{
\begin{fmfgraph}(60,40)
\fmfleft{i1}
\fmfright{o1}
\fmf{plain}{i1,o1}
\end{fmfgraph}}
\end{fmffile} = i \Delta(p)\,,
\end{equation}
where $\Delta(p)$ is the Fourier transform of $\Delta_x$, defined such that $\Delta(p) = 1/(p^2-m^2+i\epsilon)$ for $\Delta_x^{-1} = \square+m^2$.
Alternatively, one could treat terms in $G[\phi]$ that are linear or constant in $\phi$ to all orders directly by resumming them into the propagator.
We will explain how field-redefinition invariance manifests for this approach in \cref{sec:Resummation} below.

The third term in \cref{eq:lagrPrime} gives rise to a $(2n)$-point vertex
\begin{equation}
\label{eq:diagram_GG}
\quad \begin{fmffile}{tree_vert2_a}
\parbox{95pt}{
\begin{fmfgraph*}(60,40)
\fmfleft{i0,i1,i2,i3,i4}
\fmfright{o0,o1,o2,o3,o4}
\fmfv{l={$G[\phi]$}}{i2}
\fmfv{l={$G[\phi]$}}{o2}
\fmf{plain,foreground=(.9,,0.1,,0.1)}{i1,v1}
\fmf{phantom,foreground=(.9,,0.1,,0.1)}{i2,v1}
\fmf{plain,foreground=(.9,,0.1,,0.1)}{i3,v1}
\fmf{dots,tension=0}{i1,i2}
\fmf{dots,tension=0}{i3,i2}
\fmf{dots,tension=0}{o1,o2}
\fmf{dots,tension=0}{o3,o2}
\fmfblob{.15w}{v1}
\fmf{plain,foreground=(.9,,0.1,,0.1)}{v1,o1}
\fmf{phantom,foreground=(.9,,0.1,,0.1)}{v1,o2}
\fmf{plain,foreground=(.9,,0.1,,0.1)}{v1,o3}
\end{fmfgraph*}}
\end{fmffile} 
= i\lambda^2 G(p_1,\ldots,p_n)\, G(p_{n+1},\ldots,p_{2n}) \,\Delta^{-1}\left(\textstyle{\sum}_{i=1}^n p_{i}\right)  \, .
\end{equation}
The case of $n=1$ corresponds to a propagator insertion.

The second term in the Lagrangian in \cref{eq:lagrPrime} gives rise to an $(n+1)$-point vertex 
\begin{equation}\label{eq:3vertex}
\begin{fmffile}{tree_vert1_1}
\parbox{80pt}{
\begin{fmfgraph*}(60,40)
\fmfleft{i0,i1,i2,i3,i4}
\fmfright{o1}
\fmf{plain,foreground=(.9,,0.1,,0.1)}{i1,v1}
\fmf{phantom,foreground=(.9,,0.1,,0.1)}{i2,v1}
\fmf{plain,foreground=(.9,,0.1,,0.1)}{i3,v1}
\fmf{dots,tension=0}{i1,i2}
\fmf{dots,tension=0}{i3,i2}
\fmfv{l={$G[\phi]$}}{i2}
\fmfv{l={$\phi$}}{o1}
\fmfblob{.15w}{v1}
\fmf{plain,left=0,label=$p_{n+1}$,tension=2}{v1,o1}
\fmf{phantom_arrow,left=0,tension=0,foreground=(.9,,0.2,,0.9)}{v1,o1}
\end{fmfgraph*}}
\end{fmffile} = i \lambda  G(p_1,\ldots,p_n)\, \Delta^{-1}(p_{n+1})\,,
\end{equation}
where $n$ is the number of legs in $G[\phi]$.
The arrow denotes the leg that is acted on by the operator $\Delta^{-1}$, and the function $G(p)$ 
encodes the Feynman rule for $G[\phi]$.
The combination of 
two such $(n+1)$-point vertices cancels against the contribution in \cref{eq:diagram_GG} for any $n$: 
\begin{align}
\begin{fmffile}{tree_forwardbackward}
\parbox{70pt}{
\begin{fmfgraph*}(70,40)
\fmfleft{i0,i1,i2,i3,i4}
\fmfright{o0,o1,o2,o3,o4}
\fmf{plain,foreground=(.9,,0.1,,0.1)}{i1,v1}
\fmf{phantom,foreground=(.9,,0.1,,0.1)}{i2,v1}
\fmf{plain,foreground=(.9,,0.1,,0.1)}{i3,v1}
\fmfblob{.125w}{v1}
\fmf{plain,left=0,tension=2}{v1,v2}
\fmf{phantom_arrow,left=0,tension=0,foreground=(.9,,0.2,,0.9)}{v1,v2}
\fmf{plain,tension=4,label=$q$}{v2,v3}
\fmf{plain,left=0,tension=2}{v4,v3}
\fmf{phantom_arrow,left=0,tension=0,foreground=(.9,,0.2,,0.9)}{v4,v3}
\fmf{dots,tension=0}{i1,i2}
\fmf{dots,tension=0}{i3,i2}
\fmf{dots,tension=0}{o1,o2}
\fmf{dots,tension=0}{o3,o2}
\fmfblob{0.125w}{v4}
\fmf{plain,foreground=(.9,,0.1,,0.1)}{v4,o1}
\fmf{phantom,foreground=(.9,,0.1,,0.1)}{v4,o2}
\fmf{plain,foreground=(.9,,0.1,,0.1)}{v4,o3}
\end{fmfgraph*}}
\end{fmffile} &= (i \lambda) G(p_1,\ldots,p_n)\Delta^{-1}\left(q\right) \left[i \Delta\left(q\right)\right]
(i\lambda) G(p_{n+1},\ldots,p_{2n}) \Delta^{-1}\left(q\right)\notag\\
&= -i\lambda^2 G(p_1,\ldots,p_n) G(p_{n+1},\ldots,p_{2n})\Delta^{-1}\left(\textstyle{\sum}_{i=1}^{n} p_i\right)\notag\\[4pt]
& = - \left(\begin{fmffile}{tree_vert2}
\parbox{60pt}{
\begin{fmfgraph}(60,40)
\fmfleft{i0,i1,i2,i3,i4}
\fmfright{o0,o1,o2,o3,o4}
\fmf{plain,foreground=(.9,,0.1,,0.1)}{i1,v1}
\fmf{phantom,foreground=(.9,,0.1,,0.1)}{i2,v1}
\fmf{plain,foreground=(.9,,0.1,,0.1)}{i3,v1}
\fmf{dots,tension=0}{i1,i2}
\fmf{dots,tension=0}{i3,i2}
\fmf{dots,tension=0}{o1,o2}
\fmf{dots,tension=0}{o3,o2}
\fmfblob{.15w}{v1}
\fmf{plain,foreground=(.9,,0.1,,0.1)}{v1,o1}
\fmf{phantom,foreground=(.9,,0.1,,0.1)}{v1,o2}
\fmf{plain,foreground=(.9,,0.1,,0.1)}{v1,o3}
\end{fmfgraph}}
\end{fmffile}\right) \,.
\label{eq:diagram_two_arrows}
\end{align}
This cancellation occurs off-shell and for completely arbitrary forms of $G[\phi]$. Of course, in the case where $G[\phi]$ is a general polynomial in $\phi$ rather than a monomial $\phi^n$, the conclusion remains unchanged.

\subsubsection*{Case (b)}
Now consider the interaction Lagrangian $\mathcal{L}_\text{int}[\phi]$.
For simplicity, we will assume that $\mathcal{L}_\text{int}[\phi]$ contains only terms with $m$ factors of $\phi$, though it generalizes easily.
$\mathcal{L}_\text{int}[\phi]$ then gives us an $m$-point vertex, indicated by a green blob:
\begin{equation}
\begin{fmffile}{tree_fphi}
\parbox{60pt}{
\begin{fmfgraph*}(60,40)
\fmfleft{i1,i2,i3,imid,i4,i5,i6}
\fmfright{o1,o2,o3,omid,o4,o5,o6}
\fmf{plain,tension=5}{imid,v1}
\fmf{dots,tension=0,right=0.1}{o2,o5}
\fmf{phantom,tension=1}{v1,o2}
\fmf{phantom,tension=1}{v1,o5}
\fmf{plain,tension=1}{v1,o2}
\fmf{phantom,tension=1}{v1,o3}
\fmf{phantom,tension=1}{v1,o4}
\fmf{plain,tension=1}{v1,o5}
\fmfv{decor.shape=circle,decor.filled=30,decor.size=10,foreground=(.1,,0.5,,0.1)}{v1}
\end{fmfgraph*}}
\end{fmffile}
= i f_\text{int}(p_1,\ldots,p_m) \, .
\end{equation}
The field redefinition generates new interactions of the form
\begin{equation}
    \label{eq:fExpand}
    \mathcal{L}_\text{int}(\phi+\lambda G[\phi]) = \mathcal{L}_\text{int}[\phi] + \lambda \mathcal{L}_\text{int}^{(1)} G[\phi] + \frac{\lambda^2}{2} \mathcal{L}_\text{int}^{(2)} G[\phi]^2  + \cdots \,,
\end{equation}
where $\mathcal{L}_\text{int}^{(n)} \equiv \frac{\delta^n}{ \delta\phi^n} \mathcal{L}_{\text{int}}[\phi]$.
The new interactions correspond to vertices where $\phi$ legs are replaced with $G[\phi]$ legs,
\begin{equation}
\label{eq:f_expandsion}
\begin{fmffile}{tree_fphiprime_0}
\parbox{60pt}{
\begin{fmfgraph*}(60,40)
\fmfleft{i1,i2,i3,imid,i4,i5,i6}
\fmfright{o1,o2,o3,omid,o4,o5,o6}
\fmf{plain,tension=5}{imid,v1}
\fmf{dots,tension=0,right=0.1}{o2,o5}
\fmf{phantom,tension=1}{v1,o2}
\fmf{phantom,tension=1}{v1,o5}
\fmf{plain,tension=1}{v1,o2}
\fmf{phantom,tension=1}{v1,o3}
\fmf{phantom,tension=1}{v1,o4}
\fmf{plain,tension=1}{v1,o5}
\fmfv{decor.shape=circle,decor.filled=30,decor.size=10,foreground=(.1,,0.5,,0.1)}{v1}
\end{fmfgraph*}}
\end{fmffile}
\quad + \quad
\begin{fmffile}{tree_fphiprime_1}
\parbox{60pt}{
\begin{fmfgraph*}(60,40)
\fmfleft{i1,i2,i3,imid,i4,i5,i6}
\fmfright{o1,o2,o3,omid,o4,o5,o6}
\fmf{plain,foreground=(.9,,0.1,,0.1),tension=1}{i2,v1}
\fmf{phantom,tension=1}{i1,v1}
\fmf{phantom,tension=1}{i3,v1}
\fmf{phantom,tension=1}{i4,v1}
\fmf{phantom,tension=1}{i5,v1}
\fmf{dots,tension=0,right=0.1}{o2,o5}
\fmf{dots,tension=0,left=0.1}{i2,i5}
\fmf{plain,foreground=(.9,,0.1,,0.1),tension=1}{i5,v1}
\fmf{phantom,tension=1}{v1,o2}
\fmf{phantom,tension=1}{v1,o5}
\fmf{plain,tension=1}{v1,o2}
\fmf{phantom,tension=1}{v1,o3}
\fmf{phantom,tension=1}{v1,o4}
\fmf{plain,tension=1}{v1,o5}
\fmfv{decor.shape=circle,decor.filled=30,decor.size=10,foreground=(.1,,0.5,,0.1)}{v1}
\end{fmfgraph*}}
\end{fmffile}
\quad +\quad
\begin{fmffile}{tree_fphiprime_2}
\parbox{60pt}{
\begin{fmfgraph*}(60,40)
\fmfleft{i1,i2,i3,imid,i4,i5,i6}
\fmfright{o1,o2,o3,omid,o4,o5,o6}
\fmf{plain,foreground=(.9,,0.1,,0.1),tension=1}{i1,v1}
\fmf{phantom,foreground=(.9,,0.1,,0.1),tension=1}{i2,v1}
\fmf{plain,foreground=(.9,,0.1,,0.1),tension=1}{i3,v1}
\fmf{plain,foreground=(.9,,0.1,,0.1),tension=1}{i4,v1}
\fmf{phantom,foreground=(.9,,0.1,,0.1),tension=1}{i5,v1}
\fmf{plain,foreground=(.9,,0.1,,0.1),tension=1}{i6,v1}
\fmf{dots,tension=0,right=0.1}{o2,o5}
\fmf{dots,tension=0,right=0.1}{i6,i5}
\fmf{dots,tension=0,right=0.1}{i5,i4}
\fmf{dots,tension=0,right=0.1}{o2,o5}
\fmf{dots,tension=0,right=0.1}{i3,i2}
\fmf{dots,tension=0,right=0.1}{i2,i1}
\fmf{plain,tension=1}{v1,o2}
\fmf{dots,tension=0,right=0.1}{o2,o5}
\fmf{plain,tension=1}{v1,o5}
\fmf{phantom,tension=1}{v1,o1}
\fmf{phantom,tension=1}{v1,o3}
\fmf{phantom,tension=1}{v1,o4}
\fmf{phantom,tension=1}{v1,o6}
\fmfv{decor.shape=circle,decor.filled=30,decor.size=10,foreground=(.1,,0.5,,0.1)}{v1}
\end{fmfgraph*}}
\end{fmffile}
\quad +\quad \cdots\,\,.
\end{equation}
Here, the first term is $G$-independent, while the second and third terms are proportional to $G$ and $G^2$, respectively.

We will show the cancellation explicitly for the simplest case where we have one factor of $G$ and $\mathcal{L}_\text{int}^{(1)}$.   
The contribution from the second diagram in \cref{eq:f_expandsion} is given by 
\begin{equation}\label{eq:fphi_ins}
\begin{fmffile}{tree_fphiprime_1_1}
\parbox{60pt}{
\begin{fmfgraph*}(60,40)
\fmfleft{i1,i2,i3,imid,i4,i5,i6}
\fmfright{o1,o2,o3,omid,o4,o5,o6}
\fmf{plain,foreground=(.9,,0.1,,0.1),tension=1}{i2,v1}
\fmf{phantom,tension=1}{i1,v1}
\fmf{phantom,tension=1}{i3,v1}
\fmf{phantom,tension=1}{i4,v1}
\fmf{phantom,tension=1}{i5,v1}
\fmf{dots,tension=0,right=0.1}{o2,o5}
\fmf{dots,tension=0,left=0.1}{i2,i5}
\fmf{plain,foreground=(.9,,0.1,,0.1),tension=1}{i5,v1}
\fmf{phantom,tension=1}{v1,o2}
\fmf{phantom,tension=1}{v1,o5}
\fmf{plain,tension=1}{v1,o2}
\fmf{phantom,tension=1}{v1,o3}
\fmf{phantom,tension=1}{v1,o4}
\fmf{plain,tension=1}{v1,o5}
\fmfv{decor.shape=circle,decor.filled=30,decor.size=10,foreground=(.1,,0.5,,0.1)}{v1}
\end{fmfgraph*}}
\end{fmffile} = i \lambda G(p_i,\ldots,p_{i+n-1})\,f_\text{int}^{(1)}\!\Big(p_1,\ldots,\textstyle{\sum}_{j=i}^{i+n-1} p_{j},\ldots,p_{n+m-1}\Big) \, ,
\end{equation}
where $G[\phi]$ has been inserted on leg $j$, and $f_\text{int}^{(1)}(p)$ is the Feynman rule from $\mathcal{L}_\text{int}^{(1)}$.
Now consider the diagram where we insert the $(n+1)$-point $G$ vertex onto leg $i$ of the $\mathcal{L}_\text{int}[\phi]$ vertex:
\begin{align}
\begin{fmffile}{tree_fphiprime_propins}
\parbox{60pt}{
\begin{fmfgraph*}(60,40)
\fmfleft{i1,i2,i3,i4,i5}
\fmfright{o1,o2,o3,o4,o5}
\fmfblob{0.15w}{v1}
\fmf{plain,foreground=(.9,,0.1,,0.1),tension=1}{i2,v1}
\fmf{phantom,foreground=(.9,,0.1,,0.1),tension=1}{i3,v1}
\fmf{plain,foreground=(.9,,0.1,,0.1),tension=1}{i4,v1}
\fmf{plain,tension=2,label=$q$}{v1,v2}
\fmf{phantom_arrow,tension=0,foreground=(.9,,0.2,,0.9)}{v1,v2}
\fmf{plain,tension=1}{v2,o2}
\fmf{phantom,tension=1}{v2,o3}
\fmf{plain,tension=1}{v2,o4}
\fmf{dots,tension=0}{i2,i3}
\fmf{dots,tension=0}{i4,i3}
\fmf{dots,tension=0}{o2,o3}
\fmf{dots,tension=0}{o4,o3}
\fmfv{decor.shape=circle,decor.filled=30,decor.size=10,foreground=(.1,,0.5,,0.1)}{v2}
\end{fmfgraph*}}
\end{fmffile}
&= (i \lambda) G(p_i,\ldots,p_{i+n-1}) \Delta^{-1}(q) \left[i\Delta(q)\right] \left(i f_\text{int}^{(1)}\!(p_1,\ldots,q,\ldots,p_{n+m-1})\right)\notag\\
&= -i\lambda G(p_i,\ldots,p_{i+n-1}) \,f_\text{int}^{(1)}\!\Big(p_1,\ldots,\textstyle{\sum}_{j=i}^{i+n-1} p_{j},\ldots,p_{n+m-1}\Big)\notag\\[4pt]
&=- \left( \begin{fmffile}{tree_fphiprime_1_tmp}
\parbox{60pt}{
\begin{fmfgraph*}(60,40)
\fmfleft{i1,i2,i3,imid,i4,i5,i6}
\fmfright{o1,o2,o3,omid,o4,o5,o6}
\fmf{plain,foreground=(.9,,0.1,,0.1),tension=1}{i2,v1}
\fmf{phantom,tension=1}{i1,v1}
\fmf{phantom,tension=1}{i3,v1}
\fmf{phantom,tension=1}{i4,v1}
\fmf{phantom,tension=1}{i5,v1}
\fmf{dots,tension=0,right=0.1}{o2,o5}
\fmf{dots,tension=0,left=0.1}{i2,i5}
\fmf{plain,foreground=(.9,,0.1,,0.1),tension=1}{i5,v1}
\fmf{phantom,tension=1}{v1,o2}
\fmf{phantom,tension=1}{v1,o5}
\fmf{plain,tension=1}{v1,o2}
\fmf{phantom,tension=1}{v1,o3}
\fmf{phantom,tension=1}{v1,o4}
\fmf{plain,tension=1}{v1,o5}
\fmfv{decor.shape=circle,decor.filled=30,decor.size=10,foreground=(.1,,0.5,,0.1)}{v1}
\end{fmfgraph*}}
\end{fmffile}\right) \,,
\end{align}
so it exactly cancels the contribution from \cref{eq:fphi_ins} fully off-shell.
Here, $f_\text{int}^{(1)}$ appears instead of $f_\text{int}$ to account for the correct symmetry factor since using $\mathcal{L}_\text{int}^{(1)}$ sums over insertions of the $(n+1)$-point vertex. 

The same cancellation takes place for the full Taylor expansion of $f_\text{int}$ in \cref{eq:fExpand}.  Additionally, this argument generalizes straightforwardly to an arbitrary number of insertions of $G[\phi]$.  This demonstrates the full cancellation for case (b).

\subsubsection*{Case (c)}

We will now explain how the cancellation involving the source term occurs.
The modified source term is given by
\begin{equation}
    \int \dd^4x\, J(x) \phi(x) \rightarrow \int \dd^4 x \,\bigg(J(x) \phi(x) + \lambda J(x) G[\phi(x)]\bigg) \, ,
\end{equation}
which give us the two vertices 
\begin{subequations}
\label{eq:FR_source}
\begin{align}
\begin{fmffile}{source}
\parbox{40pt}{
\begin{fmfgraph*}(40,40)
\fmfcmd{
    path quadrant, q[], otimes;
    quadrant = (0, 0) -- (0.5, 0) & quartercircle & (0, 0.5) -- (0, 0);
    for i=1 upto 4: q[i] = quadrant rotated (45 + 90*i); endfor
    otimes = q[1] & q[2] & q[3] & q[4] -- cycle;
}
\fmfwizard
\fmfleft{i1}
\fmfright{o1}
\fmf{plain,tension=2}{i1,o1}
\fmfv{d.sh=otimes,d.f=empty,d.si=10,label=$J$,l.d=10}{i1}
\end{fmfgraph*}}
\end{fmffile} 
&= 1\,, \\
\begin{fmffile}{source_redef}
\parbox{40pt}{
\begin{fmfgraph*}(40,40)
\fmfcmd{
    path quadrant, q[], otimes;
    quadrant = (0, 0) -- (0.5, 0) & quartercircle & (0, 0.5) -- (0, 0);
    for i=1 upto 4: q[i] = quadrant rotated (45 + 90*i); endfor
    otimes = q[1] & q[2] & q[3] & q[4] -- cycle;
}
\fmfwizard
\fmfright{i1,i2,i3,i4,i5}
\fmfleft{o1}
\fmfblob{0.15w}{o1}
\fmf{plain,foreground=(.9,,0.1,,0.1),tension=1}{i2,o1}
\fmf{phantom,foreground=(.9,,0.1,,0.1),tension=1}{i3,o1}
\fmf{plain,foreground=(.9,,0.1,,0.1),tension=1}{i4,o1}
\fmf{dots,tension=0}{i2,i3}
\fmf{dots,tension=0}{i4,i3}
\fmfv{d.sh=otimes,d.f=empty,d.si=10,label=$J$,l.d=10}{o1}
\end{fmfgraph*}}
\end{fmffile} &= \lambda G(p_1,\ldots,p_n) \,. \label{eq:new_source}
\end{align}
\end{subequations}
We also have diagrams of the form
\begin{align}
\begin{fmffile}{tree_source}
\parbox{80pt}{
\begin{fmfgraph*}(60,40)
\fmfcmd{
    path quadrant, q[], otimes;
    quadrant = (0, 0) -- (0.5, 0) & quartercircle & (0, 0.5) -- (0, 0);
    for i=1 upto 4: q[i] = quadrant rotated (45 + 90*i); endfor
    otimes = q[1] & q[2] & q[3] & q[4] -- cycle;
}
\fmfwizard
\fmfright{i1,i2,i3,i4,i5}
\fmfleft{o1}
\fmfblob{0.15w}{v1}
\fmf{plain,foreground=(.9,,0.1,,0.1),tension=1}{i2,v1}
\fmf{phantom,foreground=(.9,,0.1,,0.1),tension=1}{i3,v1}
\fmf{plain,foreground=(.9,,0.1,,0.1),tension=1}{i4,v1}
\fmf{plain,tension=2}{v1,o1}
\fmf{phantom_arrow,tension=0,foreground=(.9,,0.2,,0.9)}{v1,o1}
\fmf{dots,tension=0}{i2,i3}
\fmf{dots,tension=0}{i4,i3}
\fmfv{d.sh=otimes,d.f=empty,d.si=10,l.d=10}{o1}
\end{fmfgraph*}}
\end{fmffile} 
&= i \lambda  G(p_1,\ldots,p_n) \Delta^{-1}(p_{n+1}) \left[i\Delta(p_{n+1})\right] \notag\\
&= - \left( \ \begin{fmffile}{tree_source_redef}
\parbox{40pt}{
\begin{fmfgraph*}(40,40)
\fmfcmd{
    path quadrant, q[], otimes;
    quadrant = (0, 0) -- (0.5, 0) & quartercircle & (0, 0.5) -- (0, 0);
    for i=1 upto 4: q[i] = quadrant rotated (45 + 90*i); endfor
    otimes = q[1] & q[2] & q[3] & q[4] -- cycle;
}
\fmfwizard
\fmfright{i1,i2,i3,i4,i5}
\fmfleft{o1}
\fmfblob{0.15w}{o1}
\fmf{plain,foreground=(.9,,0.1,,0.1),tension=1}{i2,o1}
\fmf{phantom,foreground=(.9,,0.1,,0.1),tension=1}{i3,o1}
\fmf{plain,foreground=(.9,,0.1,,0.1),tension=1}{i4,o1}
\fmf{dots,tension=0}{i2,i3}
\fmf{dots,tension=0}{i4,i3}
\fmfv{d.sh=otimes,d.f=empty,d.si=10,l.d=10}{o1}
\end{fmfgraph*}}
\end{fmffile}\right) \,,
\end{align}
which clearly cancel.
This completes the argument that all new diagrams involving $(G\Delta^{-1} G)$, $\sum_{n} \frac{1}{n!}\mathcal{L}_{\text{int}}^{(n)} G^n$, or $(JG)$ are precisely canceled by diagrams that involve $(G\Delta^{-1} \phi)$. 

Depending on the specific form of $G[\phi]$, the changes in the source may often be neglected when going from off-shell correlation functions to $S$-matrix elements.
We will say more about this in \cref{sec:smatrix}.

\subsubsection*{Case (d)}
The last case comes from diagrams where the arrows flow along a closed loop:
\begin{equation}\label{eq:ring1}
\begin{fmffile}{ring}
\parbox{100pt}{
  \begin{fmfgraph*}(100,80)
\fmfleft{i0,i1,i3,i4}
\fmfright{o0,o1,o3,o4}
\fmfblob{.07w}{v1}
\fmfblob{.07w}{v2}
\fmfblob{.07w}{v3}
\fmfblob{.07w}{v4}
\fmf{dots,tension=0,left=0.1}{i0,i1}
\fmf{dots,tension=0,left=0.1}{i3,i4}
\fmf{dots,tension=0,right=0.1}{o0,o1}
\fmf{dots,tension=0,right=0.1}{o3,o4}
\fmf{plain,foreground=(.1,,0.1,,0.9)}{i0,v1}
\fmf{plain,foreground=(.1,,0.1,,0.9)}{i1,v1}
\fmf{plain,foreground=(.1,,0.1,,0.9)}{i3,v2}
\fmf{plain,foreground=(.1,,0.1,,0.9)}{i4,v2}
\fmf{plain,left=0.5,tension=1.5}{v1,v2,v3,v4,v1}
\fmf{phantom_arrow,left=0.5,tension=0,foreground=(.9,,0.2,,0.9)}{v1,v2,v3,v4,v1}
\fmf{plain,foreground=(.1,,0.1,,0.9),tension=1}{v4,o0}
\fmf{plain,foreground=(.1,,0.1,,0.9),tension=1}{v4,o1}
\fmf{plain,foreground=(.1,,0.1,,0.9),tension=1}{v3,o3}
\fmf{plain,foreground=(.1,,0.1,,0.9),tension=1}{v3,o4}
  \end{fmfgraph*}}\quad,
  \end{fmffile}
\end{equation}
where clusters of $(n-1)$ blue lines indicate an insertion of $\frac{\delta G}{\delta \phi} \phi_n$, where $\phi_n$ denotes the internal line without the arrow leaving the vertex:
\begin{equation}
\label{eq:deltaG_int}
\begin{fmffile}{deltaG_int}
  \parbox{75pt}{\begin{fmfgraph*}(60,40)
    \fmfleft{i0,i1,i2,i3,i4}
    \fmfright{o1,o2}
    \fmfv{l={$\frac{\delta G[\phi]}{\delta \phi}$}}{i2}
    \fmf{plain,foreground=(.1,,0.1,,0.9)}{i1,v1}
    \fmf{plain,foreground=(.1,,0.1,,0.9)}{i3,v1}
    \fmf{dots,tension=0,left=0.1}{i1,i3}
    \fmfblob{0.15w}{v1}
    \fmf{plain}{v1,o2}
    \fmf{phantom_arrow,foreground=(.9,,0.2,,0.9),tension=0}{v1,o2}
    \fmf{plain}{o1,v1}
    \fmfv{l={$p_n$},l.a=0}{o1}
    \fmfv{l={$p_{n+1}$},l.a=0}{o2}
  \end{fmfgraph*}}
\end{fmffile} = i \lambda G^{(1)}(p_1,\ldots,p_n) \Delta^{-1}(p_{n+1})\,.\\[1.5em] 
\end{equation}
Writing the interaction this way effectively sums over configurations for the internal line, following the same logic as in case (b).
For a general $G[\phi]$, diagrams of the form of~\cref{eq:ring1} are non-zero, and are canceled by the determinant in \cref{eq:Zj}.
The determinant can be written as an integral over auxiliary ghost fields as
\begin{equation}\label{eq:ghosts}
    \text{det}\bigg(\frac{\delta \widetilde{\phi}}{\delta \phi}\bigg) = \int\! \mathcal{D} \bar{c}\, \mathcal{D} c\, \exp\!{\left\{-i\!\int\! \dd^4 x \left( \bar{c} c + \lambda \bar{c} \frac{\delta G}{\delta \phi} c \right)\right\}} \, .
\end{equation}
Written this way, the ghost propagator is trivial,
\begin{equation}
\begin{fmffile}{prop_gh}
\parbox{60pt}{
\begin{fmfgraph}(60,40)
\fmfleft{i1}
\fmfright{o1}
\fmf{dots_arrow}{i1,o1}
\end{fmfgraph}}
\end{fmffile} = i \, ,
\end{equation}
where the arrow on the ghost propagator keeps track of ghost number and does \textit{not} indicate a $\Delta^{-1}$ insertion, in contrast to the scalar case.
The interaction term gives rise to the vertex
\begin{equation}
\begin{fmffile}{ghost_int}
  \parbox{75pt}{\begin{fmfgraph*}(60,40)
    \fmfleft{i0,i1,i2,i3,i4}
    \fmfright{o1,o2}
    \fmfv{l={$\frac{\delta G[\phi]}{\delta \phi}$}}{i2}
    \fmf{plain,foreground=(.1,,0.1,,0.9)}{i1,v1}
    \fmf{plain,foreground=(.1,,0.1,,0.9)}{i3,v1}
    \fmf{dots,tension=0,left=0.1}{i1,i3}
    \fmfblob{0.15w}{v1}
    \fmf{dots_arrow}{v1,o2}
    \fmf{dots_arrow}{o1,v1}
    \fmfv{l={$p_n$},l.a=0}{o1}
  \end{fmfgraph*}}
\end{fmffile} = i \lambda  G^{(1)}(p_1,\ldots,p_n) \,.
\end{equation}
The kinematic function in the ghost vertex is identical to the one in the vertex in \cref{eq:deltaG_int}. 
This allows us to write
\begin{equation}\label{eq:ghost_canc}
\begin{fmffile}{tree_vert1}
\parbox{60pt}{
\begin{fmfgraph*}(60,40)
\fmfleft{i0,i1,i2,i3,i4}
\fmfright{o1}
\fmf{plain}{i1,v1}
\fmf{plain,foreground=(.1,,0.1,,0.9)}{i2,v1}
\fmf{plain,foreground=(.1,,0.1,,0.9)}{i3,v1}
    \fmf{dots,tension=0,left=0.1}{i2,i3}
\fmfblob{.15w}{v1}
\fmf{plain,left=0,tension=2}{v1,o1}
\fmf{phantom_arrow,left=0,tension=0,foreground=(.9,,0.2,,0.9)}{v1,o1}
\end{fmfgraph*}}
\end{fmffile}  \ \ 
\times \ \
\begin{fmffile}{nonlocal_tmp3}
\parbox{70pt}{
\begin{fmfgraph}(60,40)
\fmfleft{i1}
\fmfright{o1}
\fmf{plain}{i1,o1}
\end{fmfgraph}}
\end{fmffile} 
= \ \ 
\begin{fmffile}{tree_vert_tmp3}
\parbox{60pt}{
\begin{fmfgraph*}(60,40)
\fmfleft{i0,i1,i2,i3,i4}
\fmfright{o1}
\fmf{dots_arrow}{i1,v1}
\fmf{plain,foreground=(.1,,0.1,,0.9)}{i2,v1}
\fmf{plain,foreground=(.1,,0.1,,0.9)}{i3,v1}
\fmf{dots,tension=0,left=0.1}{i2,i3}
\fmfblob{.15w}{v1}
\fmf{dots_arrow,left=0,tension=2}{v1,o1}
\end{fmfgraph*}}
\end{fmffile} \ \ 
\times \ \ 
\begin{fmffile}{nonlocal_gh_tmp2}
\parbox{60pt}{
\begin{fmfgraph}(60,40)
\fmfleft{i1}
\fmfright{o1}
\fmf{dots_arrow}{i1,o1}
\end{fmfgraph}}\,\,,
\end{fmffile} 
\end{equation}
which means that diagrams with an arrow flowing around a closed loop precisely cancel the ghost diagrams, due to  the ghost's additional factor of $(-1)$ that comes from the Grassmann algebra:
\begin{equation}
\begin{fmffile}{ring2}
\parbox{100pt}{
  \begin{fmfgraph*}(100,80)
\fmfleft{i0,i1,i3,i4}
\fmfright{o0,o1,o3,o4}
\fmfblob{.07w}{v1}
\fmfblob{.07w}{v2}
\fmfblob{.07w}{v3}
\fmfblob{.07w}{v4}
\fmf{dots,tension=0,left=0.1}{i0,i1}
\fmf{dots,tension=0,left=0.1}{i3,i4}
\fmf{dots,tension=0,right=0.1}{o0,o1}
\fmf{dots,tension=0,right=0.1}{o3,o4}
\fmf{plain,foreground=(.1,,0.1,,0.9)}{i0,v1}
\fmf{plain,foreground=(.1,,0.1,,0.9)}{i1,v1}
\fmf{plain,foreground=(.1,,0.1,,0.9)}{i3,v2}
\fmf{plain,foreground=(.1,,0.1,,0.9)}{i4,v2}
\fmf{plain,left=0.5,tension=1.5}{v1,v2,v3,v4,v1}
\fmf{phantom_arrow,left=0.5,tension=0,foreground=(.9,,0.2,,0.9)}{v1,v2,v3,v4,v1}
\fmf{plain,foreground=(.1,,0.1,,0.9),tension=1}{v4,o0}
\fmf{plain,foreground=(.1,,0.1,,0.9),tension=1}{v4,o1}
\fmf{plain,foreground=(.1,,0.1,,0.9),tension=1}{v3,o3}
\fmf{plain,foreground=(.1,,0.1,,0.9),tension=1}{v3,o4}
  \end{fmfgraph*}}
  \end{fmffile} \quad +\quad \begin{fmffile}{ring3}
\parbox{100pt}{
  \begin{fmfgraph*}(100,80)
\fmfleft{i0,i1,i3,i4}
\fmfright{o0,o1,o3,o4}
\fmfblob{.07w}{v1}
\fmfblob{.07w}{v2}
\fmfblob{.07w}{v3}
\fmfblob{.07w}{v4}
\fmf{dots,tension=0,left=0.1}{i0,i1}
\fmf{dots,tension=0,left=0.1}{i3,i4}
\fmf{dots,tension=0,right=0.1}{o0,o1}
\fmf{dots,tension=0,right=0.1}{o3,o4}
\fmf{plain,foreground=(.1,,0.1,,0.9)}{i0,v1}
\fmf{plain,foreground=(.1,,0.1,,0.9)}{i1,v1}
\fmf{plain,foreground=(.1,,0.1,,0.9)}{i3,v2}
\fmf{plain,foreground=(.1,,0.1,,0.9)}{i4,v2}
\fmf{dots_arrow,left=0.5,tension=1.5}{v1,v2,v3,v4,v1}
\fmf{plain,foreground=(.1,,0.1,,0.9),tension=1}{v4,o0}
\fmf{plain,foreground=(.1,,0.1,,0.9),tension=1}{v4,o1}
\fmf{plain,foreground=(.1,,0.1,,0.9),tension=1}{v3,o3}
\fmf{plain,foreground=(.1,,0.1,,0.9),tension=1}{v3,o4}
  \end{fmfgraph*}}
  \end{fmffile} \,\,= 0\,.
\end{equation}
Since this argument is once again fully off-shell, this shows that all diagrams induced by the field redefinition cancel among themselves to all loop orders for an arbitrary number of external legs.

\subsubsection*{When Do Ghosts Matter?}
While the ghosts are necessary for a general field redefinition to enforce \cref{eq:mainPoint}, they rarely make an appearance in familiar applications.
Since ghosts only appear at loop-level, they may be dropped for any tree-level calculation.
Beyond tree level, ghosts may still be ignored for a large class of redefinitions when using dimensional regularization.
Consider the combination appearing in \cref{eq:ghost_canc}: 
\begin{equation}
\begin{fmffile}{tree_vert_tmp3}
\parbox{60pt}{
\begin{fmfgraph*}(60,40)
\fmfleft{i0,i1,i2,i3,i4}
\fmfright{o1}
\fmf{dots_arrow}{i1,v1}
\fmf{plain,foreground=(.1,,0.1,,0.9)}{i2,v1}
\fmf{plain,foreground=(.1,,0.1,,0.9)}{i3,v1}
\fmf{dots,tension=0,left=0.1}{i2,i3}
\fmfblob{.15w}{v1}
\fmf{dots_arrow,left=0,tension=2}{v1,o1}
\end{fmfgraph*}}
\end{fmffile} \ \ 
\times \ \ 
\begin{fmffile}{nonlocal_gh_tmp2}
\parbox{60pt}{
\begin{fmfgraph}(60,40)
\fmfleft{i1}
\fmfright{o1}
\fmf{dots_arrow}{i1,o1}
\end{fmfgraph}}
\end{fmffile}  = -\lambda G^{(1)}(p_1,\ldots,p_n) \, .
\end{equation}
When $G[\phi]$ is local,  $G^{(1)}(p)$ is a polynomial in momenta. 
Since the loops built out of this object have no poles, the loop integral is just an integral over a polynomial of momenta, which vanishes in dimensional regularization. 
This is typical for field redefinitions commonly used in EFTs~\cite{Arzt:1993gz,Manohar:2018aog,Criado:2018sdb}.\footnote{Another way to phrase this is that for a local EFT field redefinition of the form $\phi \rightarrow \phi + \frac{1}{\Lambda^2} G[\phi]$ with $\Lambda$ the EFT scale, the ghosts gain a mass proportional to $\Lambda$ when canonically normalized.
After expanding the ghost propagator in powers of $\Lambda$ for a consistent power counting, any ghost loop becomes a polynomial in momenta and vanishes in dim reg.
}
It should be emphasized that these ghost loops are not zero in general, for example when working with a hard momentum cutoff regulator.

The other case where they may be ignored is when $G[\phi]$ is independent of or proportional to $\phi$, i.e., $G[\phi] = A \phi + B$.
If $A$ is nonlocal then the ghost loops may be non-zero.
However, the ghosts have no interactions since $\frac{\delta G}{\delta \phi}$ is $\phi$-independent, and so they do not contribute to any correlation function.\footnote{With the important exception of the chiral anomaly for fermions~\cite{Fujikawa:1979ay}.}
In summary, when using dimensional regularization, ghosts are only relevant for nonlocal field redefinitions that are polynomials in $\phi$ with degree $n\geq2$ when going beyond tree level.  We will show this explicitly by working through a simple example in \cref{sec:RelExample} below.

\subsection{Generalization to Theories with Fermions and Gauge Bosons}
The proof can be extended to a general field theory. If we have an arbitrary number of scalars in our theory, then we can group them in multiplets $\phi^{i}$. The propagator and all interaction vertices now carry these indices. However, the proof of field-redefinition invariance proceeds as before with the appropriate contraction of the scalar indices.

For fermions, one additional subtlety arises from the ghosts. 
In \cref{eq:Zj}, we obtain determinants since the particle $\phi$ is bosonic, while for fermions we get inverse determinants.
These can still be treated using ghosts in the usual way; however, to get an inverse determinant, the ghosts are no longer Grassmann valued.
The $(-1)$ from the closed fermion loop then cancels the bosonic ghost loop, and everything proceeds as usual.

In the case of gauge bosons, gauge fixing presents a potential complication.
This is discussed nicely in Ref.~\cite{Arzt:1993gz} in the context of EFTs, though much of the discussion holds more generally.
The gauge-fixing part of the Lagrangian is given by $\mathcal{L}_\text{GF} + \mathcal{L}_\text{FP}$, where the first is the gauge-fixing term and the second is the corresponding Faddeev-Popov ghosts. 
In the case that gauge fixing is done before the field redefinition, the change in these two pieces cancel between each other exactly.\footnote{This assumes the ghosts in $\mathcal{L}_\text{FP}$ are different ghosts from those from the redefinition we discussed above. This is necessary for the new terms in $\mathcal{L}_\text{GF}$ and $\mathcal{L}_\text{FP}$ to cancel independently of the rest of the Lagrangian.}
When it is done afterward, the familiar procedure can be used to choose $\mathcal{L}_\text{GF}$ and $\mathcal{L}_\text{FP}$, although the interpretation of the choice of gauge may be different.

Also, note that relativity did not play a role in the proof. The cancellation occurred between different diagrams because the propagator $\Delta$ was canceled by the inverse propagator in the vertex $G\Delta^{-1}\phi$. For nonrelativistic theories, the exact same cancellation will take place. Therefore, we have proved the invariance under field redefinitions for correlation functions and scattering amplitudes in relativistic and nonrelativistic theories.

\subsection{Resumming Modifications into the Propagator}\label{sec:Resummation}
While the above proof works for arbitrary field redefinitions, it assumes that the field redefinition is not resummed into the definition of the propagator and is always treated perturbatively in $\lambda$. We will now relax this assumption. The two cases where this is relevant are transformations that are either linear in or independent of $\phi$. One should keep in mind that the field redefinitions discussed in this section can alter the behaviour of the field at infinity, making their non-perturbative validity unclear despite the cancellations we demonstrate here.

\subsubsection{Linear transformations}\label{sec:linear}
Here we consider the case of a linear redefinition $\phi \rightarrow A_x \phi$, where $A_x$ is a $\phi$-independent (local or nonlocal) function of spacetime derivatives or spacetime coordinates in certain cases. 
We assume the same conditions on $A_x$ as for $G$, $\Delta^{-1}$, and $\mathcal{L}_\text{int}$ in Section~\ref{sec:perturbative}: namely, that it has a Fourier transform and the resulting Feynman rules are independent of spacetime coordinates.\footnote{
Coordinate dependent nonlocal redefinitions such as $Q(x)\rightarrow e^{-i v\cdot  x} \psi(x)$ are commonplace in HQET. 
In these cases, all Feynman rules have both a quark and an antiquark, causing the exponential factors to cancel and preserving the momentum conserving delta function.}
The Lagrangian and source term become
\begin{equation}
    \mathcal{L} + J\phi \rightarrow -\frac{1}{2}( A_x \phi) \Delta_x^{-1} (A_x\phi) + \mathcal{L}_\text{int}[A_x \phi]+ J A_x \phi \, .
\end{equation} 
Resumming this interaction into the propagator amounts to defining a new propagator 
\begin{align}
\Delta_{\text{new},x}^{-1} =  \overleftarrow{A}_x\s \Delta_x^{-1} \,\overrightarrow{\!A}_x\,,
\end{align}
where the arrow denotes the direction $A_x$ acts.  With this definition, the Lagrangian is
\begin{equation}
    \mathcal{L} = -\frac{1}{2}\phi \Delta_{\text{new},x}^{-1}  \phi + \mathcal{L}_{\text{int}}[A_x \phi]\, .
\end{equation} 
Diagrammatically, the Feynman rule for the propagator becomes 
\begin{equation}
\begin{fmffile}{prop_modified}
\parbox{60pt}{
\begin{fmfgraph}(60,40)
\fmfleft{i1}
\fmfright{o1}
\fmf{dashes,foreground=(.9,,0.4,,0.1)}{i1,o1}
\end{fmfgraph}}
\end{fmffile} = i \Delta_\text{new}(p) = \frac{i \Delta(p)}{\left[A(p)\right]^2} \, .
\end{equation}
The Feynman rule for the interaction Lagrangian $\mathcal{L}_\text{int}[\phi]$ now comes with additional factors of $A(p)$:
\begin{equation}
\begin{fmffile}{tree_fphi_modified}
\parbox{60pt}{
\begin{fmfgraph*}(60,40)
\fmfleft{i1,i2,i3,imid,i4,i5,i6}
\fmfright{o1,o2,o3,omid,o4,o5,o6}
\fmf{dashes,tension=1,foreground=(.9,,0.4,,0.1)}{i2,v1}
\fmf{phantom,tension=1}{i1,v1}
\fmf{phantom,tension=1}{i3,v1}
\fmf{phantom,tension=1}{i4,v1}
\fmf{phantom,tension=1}{i5,v1}
\fmf{dots,tension=0,right=0.1}{o2,o5}
\fmf{dashes,tension=1,foreground=(.9,,0.4,,0.1)}{i5,v1}
\fmf{phantom,tension=1}{v1,o2}
\fmf{phantom,tension=1}{v1,o5}
\fmf{dashes,tension=1,foreground=(.9,,0.4,,0.1)}{o2,v1}
\fmf{phantom,tension=1}{v1,o3}
\fmf{phantom,tension=1}{v1,o4}
\fmf{dashes,tension=1,foreground=(.9,,0.4,,0.1)}{o5,v1}
\fmfv{decor.shape=circle,decor.filled=30,decor.size=10,foreground=(.1,,0.5,,0.1)}{v1}
\end{fmfgraph*}}
\end{fmffile} = i f(p_1,\ldots,p_m) \prod_{i=1}^m A(p_i) \, ,
\end{equation}
and the source terms also carry the same factor,
\begin{equation}
\begin{fmffile}{tree_source_modified}
\parbox{40pt}{
\begin{fmfgraph*}(40,40)
\fmfcmd{
    path quadrant, q[], otimes;
    quadrant = (0, 0) -- (0.5, 0) & quartercircle & (0, 0.5) -- (0, 0);
    for i=1 upto 4: q[i] = quadrant rotated (45 + 90*i); endfor
    otimes = q[1] & q[2] & q[3] & q[4] -- cycle;
}
\fmfwizard
\fmfleft{i1}
\fmfright{o1}
\fmf{dashes,foreground=(.9,,0.4,,0.1)}{i1,o1}
\fmfv{d.sh=otimes,d.f=empty,d.si=10,label=$J$,l.d=10}{i1}
\end{fmfgraph*}}
\end{fmffile} 
= A(p) \, ,
\end{equation}
such that the original theory is recovered when including sources on external legs,
\begin{equation}
\begin{fmffile}{tree_fphi_modified_2}
\parbox{70pt}{
\begin{fmfgraph*}(60,40)
\fmfcmd{
    path quadrant, q[], otimes;
    quadrant = (0, 0) -- (0.5, 0) & quartercircle & (0, 0.5) -- (0, 0);
    for i=1 upto 4: q[i] = quadrant rotated (45 + 90*i); endfor
    otimes = q[1] & q[2] & q[3] & q[4] -- cycle;
}
\fmfwizard
\fmfleft{i1,i2,i3,imid,i4,i5,i6}
\fmfright{o1,o2,o3,omid,o4,o5,o6}
\fmf{dashes,tension=1,foreground=(.9,,0.4,,0.1)}{i2,v1}
\fmf{phantom,tension=1}{i1,v1}
\fmf{phantom,tension=1}{i3,v1}
\fmf{phantom,tension=1}{i4,v1}
\fmf{phantom,tension=1}{i5,v1}
\fmf{dots,tension=0,right=0.1}{o2,o5}
\fmf{dashes,tension=1,foreground=(.9,,0.4,,0.1)}{i5,v1}
\fmf{phantom,tension=1}{v1,o2}
\fmf{phantom,tension=1}{v1,o5}
\fmf{dashes,tension=1,foreground=(.9,,0.4,,0.1)}{v1,o2}
\fmf{phantom,tension=1}{v1,o3}
\fmf{phantom,tension=1}{v1,o4}
\fmf{dashes,tension=1,foreground=(.9,,0.4,,0.1)}{v1,o5}
\fmfv{decor.shape=circle,decor.filled=30,decor.size=10,foreground=(.1,,0.5,,0.1)}{v1}
\fmfv{d.sh=otimes,d.f=empty,d.si=7}{i2}
\fmfv{d.sh=otimes,d.f=empty,d.si=7}{i5}
\fmfv{d.sh=otimes,d.f=empty,d.si=7}{o2}
\fmfv{d.sh=otimes,d.f=empty,d.si=7}{o5}
\end{fmfgraph*}}
\end{fmffile}  = \quad
\begin{fmffile}{tree_fphi_2}
\parbox{70pt}{
\begin{fmfgraph*}(60,40)
\fmfcmd{
    path quadrant, q[], otimes;
    quadrant = (0, 0) -- (0.5, 0) & quartercircle & (0, 0.5) -- (0, 0);
    for i=1 upto 4: q[i] = quadrant rotated (45 + 90*i); endfor
    otimes = q[1] & q[2] & q[3] & q[4] -- cycle;
}
\fmfwizard
\fmfleft{i1,i2,i3,imid,i4,i5,i6}
\fmfright{o1,o2,o3,omid,o4,o5,o6}
\fmf{plain,tension=1}{i2,v1}
\fmf{phantom,tension=1}{i1,v1}
\fmf{phantom,tension=1}{i3,v1}
\fmf{phantom,tension=1}{i4,v1}
\fmf{phantom,tension=1}{i5,v1}
\fmf{dots,tension=0,right=0.1}{o2,o5}
\fmf{plain,tension=1}{i5,v1}
\fmf{phantom,tension=1}{v1,o2}
\fmf{phantom,tension=1}{v1,o5}
\fmf{plain,tension=1}{v1,o2}
\fmf{phantom,tension=1}{v1,o3}
\fmf{phantom,tension=1}{v1,o4}
\fmf{plain,tension=1}{v1,o5}
\fmfv{decor.shape=circle,decor.filled=30,decor.size=10,foreground=(.1,,0.5,,0.1)}{v1}
\fmfv{d.sh=otimes,d.f=empty,d.si=7}{i2}
\fmfv{d.sh=otimes,d.f=empty,d.si=7}{i5}
\fmfv{d.sh=otimes,d.f=empty,d.si=7}{o2}
\fmfv{d.sh=otimes,d.f=empty,d.si=7}{o5}
\end{fmfgraph*}}\,,
\end{fmffile} 
\end{equation}
where solid black lines are the propagators for $\phi$ in the original theory.
For internal propagators, the two $\mathcal{L}_\text{int}[\phi]$ vertices give the two necessary factors of $A(p)$ to recover the original theory:
\begin{equation}
\begin{fmffile}{tree_twof_modifiedprop}
\parbox{70pt}{
\begin{fmfgraph*}(60,40)
\fmfleft{i1,i2,i3,i4,i5}
\fmfright{o1,o2,o3,o4,o5}
\fmfv{decor.shape=circle,decor.filled=30,decor.size=10,foreground=(.1,,0.5,,0.1)}{v1}
\fmf{plain,tension=1}{i2,v1}
\fmf{phantom,foreground=(.9,,0.4,,0.1),tension=1}{i3,v1}
\fmf{plain,tension=1}{i4,v1}
\fmf{dashes,tension=1,label=$q$,foreground=(.9,,0.4,,0.1)}{v1,v2}
\fmf{plain,tension=1}{v2,o2}
\fmf{phantom,tension=1,foreground=(.9,,0.4,,0.1)}{v2,o3}
\fmf{plain,tension=1}{v2,o4}
\fmf{dots,tension=0}{i2,i3}
\fmf{dots,tension=0}{i4,i3}
\fmf{dots,tension=0}{o2,o3}
\fmf{dots,tension=0}{o4,o3}
\fmfv{decor.shape=circle,decor.filled=30,decor.size=10,foreground=(.1,,0.5,,0.1)}{v2}
\end{fmfgraph*}}
\end{fmffile} = \quad \begin{fmffile}{tree_twof_modifiedprop_2}
\parbox{70pt}{
\begin{fmfgraph*}(60,40)
\fmfleft{i1,i2,i3,i4,i5}
\fmfright{o1,o2,o3,o4,o5}
\fmfv{decor.shape=circle,decor.filled=30,decor.size=10,foreground=(.1,,0.5,,0.1)}{v1}
\fmf{plain,tension=1}{i2,v1}
\fmf{phantom,foreground=(.9,,0.4,,0.1),tension=1}{i3,v1}
\fmf{plain,tension=1}{i4,v1}
\fmf{plain,tension=1,label=$q$}{v1,v2}
\fmf{plain,tension=1}{v2,o2}
\fmf{phantom,tension=1,foreground=(.9,,0.4,,0.1)}{v2,o3}
\fmf{plain,tension=1}{v2,o4}
\fmf{dots,tension=0}{i2,i3}
\fmf{dots,tension=0}{i4,i3}
\fmf{dots,tension=0}{o2,o3}
\fmf{dots,tension=0}{o4,o3}
\fmfv{decor.shape=circle,decor.filled=30,decor.size=10,foreground=(.1,,0.5,,0.1)}{v2}
\end{fmfgraph*}}\,.
\end{fmffile}
\end{equation}
This establishes invariance under the transformation $\phi \rightarrow A_x \phi$. 

One may wonder about the interplay of a linear transformation with a general transformation $\lambda G[\phi]$ as discussed in the previous section. 
The easiest way to see invariance under such a combination is to notice that 
\begin{align}
    \phi &\rightarrow A_x \phi +\lambda  G[\phi] \, ,
\end{align}
is equivalent to
\begin{align}
   \phi &\rightarrow A_x \phi \, , \qquad\text{then}\qquad \phi \rightarrow \phi +\lambda  A_x^{-1} G\left[\phi \right] \, . 
\end{align}
Since $G[\phi]$ was never required to be local, showing invariance under $\phi \rightarrow \phi + \lambda  G[\phi]$ already proves invariance under $\phi \rightarrow \phi +\lambda  A_x^{-1} G\left[\phi\right]$.
However, this has an interesting consequence.
Even if both $G[\phi]$ and $A_x \phi$ are strictly local transformations, ghosts are still necessary in dimensional regularization if $A_x$ contains derivatives and was resummed into the propagator since $A_x^{-1} G\left[\phi \right]$ is \textit{not} local.
Arguably the simplest example of this is in Appendix A of Ref.~\cite{Criado:2018sdb},\footnote{Explicitly, in Ref.~\cite{Criado:2018sdb} they consider a real scalar field with $\Delta_x^{-1} = \square$ and $\mathcal{L}_{\text{int}}[\phi] = 0$. 
Their field redefinition in our notation is $A_x = 1+(1/m^2)\square \phi$ and $\lambda G[\phi] = (1/m^2) g \phi^3$.} although here we see that it is completely generic.

\subsubsection{Shift Transformations}
\label{sec:FieldIndTransforms}
Now we turn to the case where the transformations are independent of the field $\phi$ entirely, $\phi(x) \rightarrow \phi(x) + \epsilon$. Here we consider $\epsilon$ to be a constant shift, i.e. independent of spacetime.\footnote{
Considering a fully general coordinate dependent function $\epsilon(x)$ generically introduces integrals over additional momenta. 
If the momentum conserving delta function remains in the Feynman rules, then the resummation may be performed in the same way as in the constant case.
Since performing the resummation and demonstrating equivalence is non-trivial even when $\epsilon$ is a constant, we do not attempt the more general case.}  This field redefinition will be used below to derive the Schwinger-Dyson equations in \cref{sec:SDEq} (although we will not need the resummed version in that section).
Some of the terms proportional to $\epsilon$ can in principle be absorbed into the propagator.  In this section, we show how to implement this resummation self-consistently.

The new Lagrangian after the field redefinition is
\begin{equation}
    \mathcal{L} = -\frac{1}{2}\phi \Delta_x^{-1} \phi -\epsilon \Delta_x^{-1} \phi  -\frac{1}{2} \epsilon \Delta_x^{-1}\epsilon +\mathcal{L}_\text{int}[\phi] + \sum_n \frac{1}{n!} \mathcal{L}_\text{int}^{(n)} \epsilon^n \,. 
\end{equation}
The term $\epsilon\Delta^{-1} \epsilon$ is $\phi\s$-independent and can be discarded. 
Which of the $\epsilon$-dependent terms can be absorbed into $\Delta_x$ depends on the form of $\mathcal{L}_\text{int}[\phi]$.
For simplicity, we specialize to the case of $\mathcal{L}_\text{int}[\phi] = -(g/3!)\phi^3$.
After shifting the field, we get
\begin{equation}
    \frac{1}{6}\s g\s \phi^3 \rightarrow \frac{1}{6}\s g\s \phi^3 + \frac{1}{2}\s g\s \epsilon\s \phi^2 + \frac{1}{2}\s  g\s \epsilon^2\phi + \frac{1}{6}\s g\s \epsilon^3\,,
\end{equation}
where $\epsilon^3$ is $\phi\s$-independent and can therefore be dropped.
The term $(g/2)\s\epsilon\s\phi^2$ can be absorbed into the definition of the propagator,
\begin{equation}
    \Delta_{\text{new},x}^{-1} =\Delta_x^{-1} +g\s\epsilon\,,
\end{equation}
so that the Lagrangian becomes 
\begin{equation}
    \label{eq:LagrPhi3Epsilon}
    \mathcal{L} = -\frac{1}{2}\phi \Delta^{-1}_{\text{new},x} \phi - \epsilon\left(\Delta^{-1}_{x} + \frac{1}{2}g\s\epsilon\right) \phi-\frac{1}{6}\s g\s\phi^3 \, .
\end{equation}
Compared to the original theory, we have a modified propagator and a new linear term.
Invariance under this field redefinition is not immediately manifest when this $\epsilon$-dependence is resummed into the propagator. We now show how the final result is in fact $\epsilon$-independent, as it must be.

First, we must check that the field redefinition has not induced a nonzero one-point function. There is now an infinite set of diagrams which contribute to the one-point function. In addition, we have a new constant coupling to the source, $\epsilon J$. After carefully resumming the infinite set of tadpole diagrams, we find that it precisely cancels the contribution from the source: 
\begin{align}
\begin{fmffile}{tadpole_resummed}
\parbox{50pt}{
\begin{fmfgraph*}(40,20)
    \fmfleft{i1} 
    \fmfright{o1}
    \fmf{plain}{i1,o1}
    \fmfv{label=\rotatebox{-90}{{$\Ground$}},l.d=0,l.a=180,foreground=(.9,,0.1,,0.1)}{i1}
  \end{fmfgraph*}}
  \end{fmffile}
  &= \quad
  \begin{fmffile}{tadpoles}
  \parbox{50pt}{
  \begin{fmfgraph*}(40,20)
    \fmfleft{i1} 
    \fmfright{o1}
    \fmf{plain}{i1,o1}
    \fmfv{decor.shape=circle,decor.filled=30,decor.size=7,l.d=9,l.a=90}{i1}
  \end{fmfgraph*}}
  + \quad \frac{1}{2}\,\times\,
 \parbox{60pt}{\begin{fmfgraph*}(50,20)
    \fmfcmd{
    path quadrant, q[], otimes;
    quadrant = (0, 0) -- (0.5, 0) & quartercircle & (0, 0.5) -- (0, 0);
    for i=1 upto 4: q[i] = quadrant rotated (45 + 90*i); endfor
    otimes = q[1] & q[2] & q[3] & q[4] -- cycle;
    }
    \fmfwizard
    \fmfleft{i1,i2} 
    \fmfright{o1}
    \fmf{plain}{i1,v1}
    \fmf{plain}{i2,v1}
    \fmf{plain,tension=2}{v1,o1}
    \fmfv{decor.shape=circle,decor.filled=30,decor.size=7}{i2}
    \fmfv{decor.shape=circle,decor.filled=30,decor.size=7}{i1}
  \end{fmfgraph*}} 
  + \quad \cdots
\end{fmffile} \nonumber\\[10pt]
    &= \left(- \epsilon\frac{\Delta^{-1}(p) - g\s\epsilon/2}{\Delta^{-1}(p) - g\s\epsilon} \right) + \frac{1}{2} \left(- \epsilon\frac{\Delta^{-1}(p) - g\s\epsilon/2}{\Delta^{-1}(p) - g\s\epsilon} \right)^2 \left( \frac{g}{\Delta^{-1}(p) - g\s\epsilon}\right) + \cdots 
    \nonumber \\[7pt] 
    &= \sum_{n=0}^{\infty} 2^{-n} c_{n} \left(- \epsilon\frac{\Delta^{-1}(p) - g\epsilon/2}{\Delta^{-1}(p) - g\epsilon} \right)^{n+1} \left( \frac{ g}{\Delta^{-1}(p) - g\epsilon}\right)^{n} = - \epsilon \,,
\end{align}
where the notation on the left-hand side denotes the resummed tadpole, the gray blobs indicate factors of $i\epsilon [\Delta^{-1}(p) - g\s \epsilon/2]$ from the linear term in the Lagrangian in \cref{eq:LagrPhi3Epsilon}, and $c_{n} = \tfrac{1}{n+1}\binom{2n}{n}$ are the Catalan numbers.

The modified propagator from the Lagrangian in \cref{eq:LagrPhi3Epsilon} is not the full two-point function of this theory. We also need to account for the diagrams that contribute to the two-point function that involve tadpole insertions. We can write the all-order expression for the modified propagator as 
\begin{align}
    \label{eq:DeltaNew}
    i\Delta_{\rm full}(p) &=
\begin{fmffile}{twopoint_tadpoles}
\begin{fmfgraph*}(70,20)
    \fmfleft{i1,i2} 
    \fmfright{o1,o2}
    \fmf{plain}{i1,v1,o1}
    \fmf{phantom}{i3,v3,o3}
    \fmffreeze
    \fmf{phantom}{v3,v2,v1}
  \end{fmfgraph*}
  +
  \begin{fmfgraph*}(70,20)
    \fmfcmd{
    path quadrant, q[], otimes;
    quadrant = (0, 0) -- (0.5, 0) & quartercircle & (0, 0.5) -- (0, 0);
    for i=1 upto 4: q[i] = quadrant rotated (45 + 90*i); endfor
    otimes = q[1] & q[2] & q[3] & q[4] -- cycle;
    }
    \fmfwizard
    \fmfleft{i1,i3} 
    \fmfright{o1,o2,o3}
    \fmf{plain}{i1,v1,o1}
    \fmf{phantom}{i3,v3,o3}
    \fmffreeze
    \fmf{plain}{v3,v2,v1}
    \fmfv{label=\rotatebox{180}{{$\Ground$}},l.d=0,l.a=180,foreground=(.9,,0.1,,0.1)}{v3}
  \end{fmfgraph*}
  + \quad 
  \begin{fmfgraph*}(70,20)
    \fmfcmd{
    path quadrant, q[], otimes;
    quadrant = (0, 0) -- (0.5, 0) & quartercircle & (0, 0.5) -- (0, 0);
    for i=1 upto 4: q[i] = quadrant rotated (45 + 90*i); endfor
    otimes = q[1] & q[2] & q[3] & q[4] -- cycle;
    }
    \fmfwizard
    \fmfstraight
    \fmfleft{i1,i2}
    \fmfright{o1,o2}
    \fmf{plain}{i1,t1}
    \fmf{phantom}{t2,i2}
    \fmf{phantom,tension=0.92}{t1,t4}
    \fmf{phantom,tension=0.92}{t2,t3}
    \fmf{plain,tension=1}{t4,o1}
    \fmf{phantom,tension=1}{o2,t3}
    \fmf{plain,tension=0}{t1,t2}
    \fmf{plain,tension=0}{t1,t4}
    \fmf{plain,tension=0}{t3,t4}
    \fmfv{label=\rotatebox{180}{{$\Ground$}},l.d=0,l.a=180,foreground=(.9,,0.1,,0.1)}{t2}
    \fmfv{label=\rotatebox{180}{{$\Ground$}},l.d=0,l.a=180,foreground=(.9,,0.1,,0.1)}{t3}
  \end{fmfgraph*}
  \quad 
  + \quad \cdots
\end{fmffile}      \nonumber \\[10pt]
    &= \frac{i}{\Delta^{-1}(p) - g\s\epsilon} +  \frac{i}{\Delta^{-1}(p) - g\s\epsilon} ( i g\s \epsilon) \frac{i}{\Delta^{-1}(p) - g\s\epsilon} + \cdots \nonumber \\[7pt]
    &= \sum_{n=0}^{\infty} \left( \frac{i}{\Delta^{-1}(p) - g\s\epsilon} \right)^{n+1}\left(i g\s \epsilon \right)^{n} = i \Delta(p) \,.
\end{align}
After resumming the tadpole contributions, the full propagator is simply the propagator in the original Lagrangian. There is an all-order cancellation between the new quadratic terms in the Lagrangian and the tadpole contributions. The cubic vertex is invariant under this field redefinition. We can use the cubic vertex and the full propagator to see that all correlation functions and $S$-matrix elements, which we discuss in \cref{sec:Implications}, are invariant, as long as we account for all tadpole corrections.

Next, we generalize the discussion to a theory with a quartic interaction, which will allow us to highlight a new feature that emerges for theories with higher-point interactions. Consider the Lagrangian
\begin{align}
    \Lagr = - \frac{1}{2} \phi \Delta_x^{-1} \phi - \frac{1}{6}\s g_{3} \phi^{3} - \frac{1}{24}\s g_{4} \phi^{4} \,.
\end{align}
After the shift of the scalar field, the Lagrangian takes the form
\begin{align}
    \Lagr &= - \frac{1}{2} \phi \left(\Delta_x^{-1} +  g_3 \epsilon + \frac{g_4 \epsilon^2}{2} \right) \phi - \epsilon \left( \Delta_{x}^{-1} + \frac{g_3 \epsilon}{2} + \frac{g_4 \epsilon^2}{6} \right) \phi\nn[5pt] 
    &\hspace{12pt}- \frac{1}{6}( g_{3} +  g_{4} \epsilon ) \phi^{3} - \frac{1}{24}g_{4} \phi^{4} \,.
\end{align}
To combat the plethora of diagrams induced by this shift, we make a simplifying choice for the shift parameter:
\begin{align}
    \epsilon = -g_3/g_4 \,.
\end{align} 
The reason this value of $\epsilon$ is useful is that it sets the cubic interactions in the redefined Lagrangian to zero.  This naively introduces a puzzle, since there are cubic interactions in the original theory.  We will see how the original cubic vertex is recovered due to including tadpole insertions in the redefined theory.

First, we check that the one-point function is zero:
\begin{align}
\begin{fmffile}{tadpole_resummed}
\parbox{50pt}{
\begin{fmfgraph*}(40,20)
    \fmfleft{i1} 
    \fmfright{o1}
    \fmf{plain}{i1,o1}
    \fmfv{label=\rotatebox{-90}{{$\Ground$}},l.d=0,l.a=180,foreground=(.9,,0.1,,0.1)}{i1}
  \end{fmfgraph*}}
  \end{fmffile}
    =& \left(- \epsilon\frac{\Delta^{-1}(p) -g_3\epsilon/2 - g_4\epsilon^2/6}{\Delta^{-1}(p) -  g_3 \epsilon - g_4\epsilon^2/2} \right) 
    \nonumber \\[6pt]
    &+ \frac{1}{6} \left(- \epsilon\frac{\Delta^{-1}(p) -g_3\epsilon/2 - g_4\epsilon^2/6}{\Delta^{-1}(p) - g_3 \epsilon - g_4\epsilon^2/2} \right)^3 \left( \frac{ g_4}{\Delta^{-1}(p) -  g_3 \epsilon - g_4\epsilon^2/2}\right) + \cdots 
    \nonumber \\[6pt] 
    =& \sum_{n=0}^{\infty} 6^{-n} d_{n} \left(- \epsilon\frac{\Delta^{-1}(p) -g_3\epsilon/2 - g_4\epsilon^2/6}{\Delta^{-1}(p) - g_3 \epsilon - g_4\epsilon^2/2} \right)^{2n+1} \left( \frac{ g_4}{\Delta^{-1}(p) - g_3 \epsilon - g_4\epsilon^2/2}\right)^{n} \nonumber \\[4pt]
    =& - \epsilon \,,
\end{align}
where $d_{n} = \tfrac{1}{2n+1}\binom{3n}{n}$, and we used $g_3 = -  g_4 \epsilon$. This cancels the contribution from the source term, $\epsilon J$.
The two-point function is also unchanged:
\begin{align}
    i \Delta_\text{full}(p)&= \frac{i}{\Delta^{-1}(p) - g_3 \epsilon - g_4\epsilon^2/2} \nonumber \\[4pt] 
    &\hspace{12pt}+  \frac{1}{2}\frac{i}{\Delta^{-1}(p) - g_3 \epsilon - g_4\epsilon^2/2} ( -i g_4 \epsilon^2) \frac{i}{\Delta^{-1}(p) - g_3 \epsilon - g_4\epsilon^2/2} + \cdots \nonumber \\[4pt]
    &= \sum_{n=0}^{\infty} \left( \frac{i}{\Delta^{-1}(p) - g_3 \epsilon - g_4\epsilon^2/2} \right)^{n+1}\left(-i g_4 \epsilon^2/2 \right)^{n} = i \Delta(p) \,,
\end{align}
where enforced that $g_3 = - g_4 \epsilon$. Finally, the three-point function is nonzero, even though there are no cubic terms in the Lagrangian:
\begin{align}
    \begin{fmffile}{g4_g3}
\parbox{50pt}{
\begin{fmfgraph*}(40,40)
    \fmfleft{i1} 
    \fmfright{o1,o2,o3}
    \fmf{plain,tension=3}{i1,v1}
    \fmf{plain}{v1,o1}
    \fmf{plain}{v1,o2}
    \fmf{plain}{v1,o3}
    \fmfv{label=\rotatebox{-90}{{$\Ground$}},l.d=0,l.a=180}{i1}
  \end{fmfgraph*}}
  \end{fmffile} = -ig_4  (-\epsilon) = -i g_3 \,.
\end{align}

The conclusions we draw from these examples hold generally. The infinite set of tadpole diagrams sums to $-\epsilon$. Furthermore, when tadpole corrections are included, all $n$-point functions in the theory after the field redefinition are the same as in the original field basis. This leads us to a statement that might be surprising at first sight; we can compute $S$-matrix elements from anywhere in field space, not just at the minimum (or extrema) of the potential. The price to pay to work away from the minimum of the potential is the inclusion of tadpole corrections.\footnote{Of course, if one does not include the tadpole corrections, then one would not get the correct result for the $S$-matrix elements away from the extrema of the potential \cite{Dudas:2023vic,Dudas:2023mmr}.} We have demonstrated here how they can be systematically included to all orders. For example, using this insight we can consider novel ways of computing $S$-matrix elements in the broken electroweak theory using the theory expanding around the unbroken point in field space. We leave this for future explorations.

\section{Implications}\label{sec:Implications}

Let us now turn to some of the implications of the general proof. 
We will first discuss correlation functions before moving on to $S$-matrix elements, paying close attention to how the LSZ procedure can be connected to the field-redefinition invariance of scattering amplitudes.
We then leverage the invariance of the path integral to derive higher-order Schwinger-Dyson equations, which imply generalized soft theorems.

\subsection{Correlation Functions}

With the diagrammatic proof of the invariance of the generating functional under field redefinitions, \cref{eq:mainPoint}, we turn our focus to correlation functions. The $n$-point correlation function is obtained from the generating functional after taking derivatives with respect to the source $J$:
\begin{align}
\langle \phi(x_1) \cdots \phi(x_n) \rangle = (-i)^n\frac{\delta}{\delta J(x_1)} \cdots \frac{\delta}{\delta J(x_n)} Z_\phi[J] \big|_{J=0}\,.
\end{align}
If we then take functional derivatives of both sides of  \cref{eq:mainPoint}, we immediately have
\begin{align}
\langle \phi(x_1)\cdots \phi(x_n) \rangle =  \langle \tilde\phi(x_1)\cdots \tilde\phi(x_n) \rangle \,.
\label{eq:correlationFnsCovariance}
\end{align}
The correlation functions are invariant under field redefinitions, as long as we compare consistently. As discussed in Ref.~\cite{Criado:2024mpx}, if one changes the coupling to the source after the field redefinition, then the resulting correlation function would be different, simply because one is computing a different quantity.

\subsection{$S$-Matrix Elements}
\label{sec:smatrix}

Until this point, we have focused on the invariance of correlation functions under field redefinitions. 
It is often the case that we are actually concerned with the implications for scattering amplitudes.  The relation between correlation functions and scattering amplitudes is given by the LSZ reduction formula~\cite{Lehmann:1954rq,Lehmann:1957zz}.
For scattering amplitudes, the common lore is that field-redefinition invariance comes from the idea that the LSZ prescription is independent of what field is used. 
This lore might seem in contradiction to our demonstration that the invariance is simply a property of correlation functions directly.
This section addresses the connection between these two points of view.

For simplicity, consider a relativistic scalar field $\phi$.
If $\phi$ is an interpolating field for a single particle state $|p\rangle$ such that 
\begin{equation}
\label{eq:interpCond}
\langle 0|\phi(x)|p\rangle \neq 0\,,
\end{equation}
then its two-point function has a pole at the physical mass $p^2 = m_\text{ph}^2$:
\begin{equation}\label{eq:lsz_twopoint}
    \int \dd^4x \ e^{ipx}\langle \phi(x)\phi(0) \rangle \sim \frac{i  R}{p^2-m_\text{ph}^2+i\epsilon} + \text{finite} \, ,
\end{equation}
where $R$ is the propagator residue.
It does not matter if we use $\phi$ or $\widetilde{\phi}$ in \cref{eq:lsz_twopoint}, as long as the field redefinition still satisfies \cref{eq:interpCond}.  In practice, this requires that the field redefinition contains a linear term.

The $n$-particle scattering amplitude is then computed from a general off-shell correlation function via the LSZ reduction formula
\begin{equation}\label{eq:lsz}
    \Bigg[\prod_{i=1}^n R_i^{1/2}\Bigg]\langle p_1,\ldots,p_n|0\rangle = \Bigg[\prod_i\lim_{p_i^2 \rightarrow m_\text{ph}^2} i \int \dd^4x_i  \ e^{i p_i x_i} \big(p_i^2-m_\text{ph}^2\big)\Bigg]\langle \phi(x_1)\cdots \phi(x_n)\rangle \, ,
\end{equation}
which projects correlation functions onto the $S$-matrix by only including contributions from $\langle \phi(x_1)\ldots\phi(x_n)\rangle$ that have poles as $p_i^2\rightarrow m_\text{ph}^2$.
Since correlation functions are invariant under field redefinitions, $S$-matrix elements are as well.

One implication of the LSZ prescription is that we can ignore contributions from \cref{eq:new_source} in the case of local field redefinitions.
Using the same source coupling $J\phi$ both before and after the redefinition leads to a different correlation function  
\begin{equation}
    \langle \phi(x_1) \ldots \phi(x_n) \rangle_{\vphantom{\widetilde{\phi}}J\phi} = \langle \widetilde{\phi}(x_1) \cdots \widetilde{\phi}(x_n)\rangle_{J\widetilde{\phi}}  \neq \langle \widetilde{\phi}(x_1) \cdots \widetilde{\phi}(x_n)\rangle_{\vphantom{\widetilde{\phi}}J\phi} \,,
\end{equation}
where the subscript indicates what coupling to the source was used.
Treating field redefinitions perturbatively as in \cref{sec:perturbative}, the difference between these correlation functions come from diagrams with arrows on external legs
\begin{align}
\begin{fmffile}{tree_onshell}
\parbox{60pt}{
\begin{fmfgraph*}(50,40)
\fmfleft{i1}
\fmfright{o1}
\fmfv{decor.shape=circle,decor.filled=30,decor.size=20}{i1}
\fmf{plain,label.side=left}{i1,v1}
\fmf{phantom_arrow,label.side=left,tension=0,foreground=(.9,,0.2,,0.9)}{i1,v1}
\fmf{plain,tension=2}{v1,o1}
\end{fmfgraph*}}
\end{fmffile} \propto \,\,G(p_n) \Delta^{-1}(p_{n}) \, ,
\end{align}
where for our relativistic scalar field, $\Delta^{-1}(p_n) \sim p_n^2-m_\text{ph}^2$, and $\Delta^{-1}(p_n) \rightarrow 0$ as $p^2_n \rightarrow m_\text{ph}^2$.
As long as $G(p_n)$ is finite as $p_n^2\rightarrow m_\text{ph}^2$, this modified source does not impact the calculation of the $S$-matrix since $\Delta^{-1}$ removes a pole in $\langle \phi(x_1)\ldots \phi(x_n)\rangle$.
This is always the case for any local field redefinition.

On the other hand, if $G(p_n)$ comes from a nonlocal field redefinition, it may not be finite as $p_n^2 \rightarrow m_\text{ph}^2$. Then the change in the source cannot be ignored. The LSZ formula still holds, but we have to also include diagrams with multiple lines attached to $J$, which can contribute to scattering amplitudes for nonlocal field redefinitions.

The LSZ formula implies that linear field redefinitions in \cref{sec:linear} have an impact; the wavefunction factors $R^{1/2}$  account for the combined change in the source and the amputated external propagators. 
The wavefunction factors can even be nonlocal functions of the kinematics.  For example, this happens when working with EFTs for heavy particles \cite{Haddad:2020tvs}. 
The $R^{1/2}$ factors can also be interpreted geometrically as tetrads \cite{Cheung:2021yog}. 
As long as the interpolating-field condition is not violated,\footnote{In the notation of \cref{sec:linear}, this translates to the requirement that $\Delta_{\text{new}}(p)$ still has a simple pole as $p^2\rightarrow m_\text{ph}^2$.} the LSZ formula will hold as usual and the $S$-matrix is invariant under field redefinitions.

\subsection{Schwinger-Dyson Equations}
\label{sec:SDEq}
The field-redefinition invariance of the path integral has physical consequences, which manifest as restrictions on relations between correlation functions.  Take the following simple transformation 
\begin{align}
    \phi(x) \rightarrow \widetilde{\phi}(x) = \phi(x) + \epsilon(x) \,,
\end{align}
where $\epsilon(x)$ is a function of spacetime. The measure is invariant under this transformation, $\mathcal{D} \widetilde{\phi} = \mathcal{D} \phi$. 
As shown in \cref{sec:FieldIndTransforms} above, correlation functions computed from the path integral must be independent of $\epsilon(x)$ at each order in $\epsilon$. Plugging this transformation into \cref{eq:correlationFnsCovariance} and expanding to first order, we have
\begin{align}
    \label{eq:SD1}
    0 &= \left\langle \left[ i \!\int\! \dd^{4}x \left( \epsilon(x) \frac{\delta\Lagr}{\delta \phi(x)} - \epsilon(x) \partial_{\mu} \frac{\delta\Lagr}{\delta \partial_{\mu}\phi(x)}\right) \right] \phi(x_1) \cdots \phi(x_2) \right\rangle 
    \nn[2pt]
    &\hspace{12pt} + \sum_{i=1}^{n}\left\langle \phi(x_1) \cdots \epsilon(x_i) \cdots \phi(x_n) \right\rangle \,.
\end{align}
This is the Schwinger-Dyson equation. The first term is the insertion of the equation-of-motion operator in a correlation function and the terms on the second line are contact terms, where one field is replaced with $\epsilon(x)$. This identity is true for any function $\epsilon(x)$. 

The Schwinger-Dyson equations can be leveraged further. After transitioning from correlation functions to scattering amplitudes, the Schwinger-Dyson equations can be used to derive a universal soft theorem \cite{Cheung:2021yog}. For a scalar effective field theory, where we drop the potential $V(\phi)$ for simplicity, the soft theorem takes a geometric form
\begin{align}
    \lim_{q\rightarrow 0} A_{n+1,i_1\dots i_n i} = \nabla_{i} A_{n,i_1\dots i_n} \,,
\end{align}
where $\nabla_{i}$ is the covariant derivative in field space. 
See Ref.~\cite{Cheung:2021yog} for details and Ref.~\cite{Derda:2024jvo} for extensions of the soft theorem to theories with scalars, fermions, and gauge bosons.

Furthermore, if we expand \cref{eq:Zj} to second order in $\epsilon$, we get a new relation between correlation functions,
\begin{align}
    \label{eq:SD2}
    0 =& \frac{1}{2} \left\langle \left[i \delta^2 S \right] \phi(x_1) \cdots \phi(x_n) \right\rangle
    + \frac{1}{2} \left\langle \left[i \delta S \right] \left[i \delta S \right] \phi(x_1) \cdots \phi(x_n) \right\rangle
    \nn &
    + \sum_{i=1}^{n}\left\langle \left[i \delta S \right] \phi(x_1) \cdots \epsilon(x_i) \cdots \phi(x_n) \right\rangle 
    + \sum_{i=1}^{n} \sum_{j\neq i} \left\langle \phi(x_1) \cdots \epsilon(x_i) \cdots \epsilon(x_j) \cdots \phi(x_n) \right\rangle \,,
\end{align}
where
\begin{subequations}
\begin{align}
    \delta S =& \int \dd^{4}x \left( \epsilon(x) \frac{\delta\Lagr}{\delta \phi(x)} - \epsilon(x) \partial_{\mu} \frac{\delta\Lagr}{\delta \partial_{\mu}\phi(x)}\right) \,, \\[5pt]
    \delta^2 S =& \int \dd^{4}x \left( \epsilon(x)\epsilon(x) \frac{\delta^2\Lagr}{\delta \phi(x)^2} + 2 \epsilon(x) \partial_{\mu}\epsilon(x) \frac{\delta^2\Lagr}{\delta \phi(x)\delta \partial_{\mu}\phi(x)} 
    \right.\nn[3pt] 
    & \left.\qquad\qquad
    +\, \partial_{\mu}\epsilon(x) \partial_{\nu}\epsilon(x) \frac{\delta^2\Lagr}{\delta \partial_{\mu}\phi(x)\delta \partial_{\nu}\phi(x)} \right) \,.
\end{align}
\end{subequations}
This is a higher-order Schwinger-Dyson equation. Obviously, one can derive analogous equations by expanding \cref{eq:Zj} to any desired order in $\epsilon$.

In \cref{eq:SD1}, we noticed that the insertion of the equation-of-motion operator in a correlation function vanished up to 
% contact terms. 
terms where one field is replaced with $\epsilon(x)$.
However, from \cref{eq:SD2} we see that the insertion of two equation-of-motion operators in a correlation function does not vanish, even when we account for 
% the contact terms. 
these terms.
We also need to include the insertion of the second variation of the action.
Nevertheless, we can use \cref{eq:SD2} to derive physical consequences for scattering amplitudes. 
The first-order Schwinger-Dyson equation leads to the soft theorem for a single particle. 
The second-order Schwinger-Dyson equation gives us the soft theorem for two particles: a double soft theorem. 
Following the same derivation as in Ref.~\cite{Cheung:2021yog} but starting with \cref{eq:SD2} and dropping the potential $V(\phi)$ for simplicity, we end up with
\begin{align}
    \lim_{q_a,q_b \rightarrow 0} A_{n+2,i_1\dots i_n i_a i_b} = \nabla_{(i_a}\nabla_{i_b)} A_{n,i_1\dots i_n } + \frac{1}{2} \sum_{c\,\neq\, a,b} \frac{s_{ac} - s_{bc}}{s_{ac} + s_{bc}} R_{i_a i_b \;\;\,i_c}^{\quad\, j_c} A_{n,i_1\dots j_c \dots i_n} \,.
\end{align}
Naturally, one can generalize this to any desired order, where the higher-order Schwinger-Dyson equations lead to multiparticle soft theorems. 
These generalized geometric soft theorems are a consequence of the field-redefinition invariance of the path integral.

\section{Examples}\label{sec:Examples}

Now that we have seen the implications of field-redefinition invariance in general, we move on to specific examples to showcase some of the most important lessons from the general proof. 
We first highlight that ghosts and modified couplings to the source must be incorporated to correctly account for nonlocal field redefinitions using a massive relativistic scalar field with a $\phi^3$ interaction.
We then show two practical nonrelativistic examples. 

\subsection{A Nonlocal Field Redefinition for a Relativistic Theory}
\label{sec:RelExample}
As a concrete example, we consider a massive scalar with a $\phi^3$ coupling
\begin{equation}
    \mathcal{L} = -\frac{1}{2}\phi\left( \square + m^2 \right) \phi - \frac{g}{3!} \phi^3\, .
\end{equation}
In the notation we used in \cref{sec:perturbative}, we have $\Delta_x^{-1} = \square + m^2$ and $\mathcal{L}_\text{int} = -(g/3!) \phi^3$.
We perform a nonlocal field redefinition with $G[\phi] = \phi \square^{-1}\phi$:
\begin{align}
\phi \rightarrow \tilde\phi = \phi + \lambda\phi\square^{-1}\phi\,.
\end{align}
After the redefinition, the Lagrangian becomes
\begin{align}\label{eq:rel_ex_Lagr}
    \mathcal{L} &= -\frac{1}{2} \phi(\square + m^2) \phi - \left[\phi\frac{\lambda}{\square} \phi\right]\left(\square+m^2\right)\phi
    - \frac{1}{2}\left[\phi\frac{\lambda}{\square} \phi\right]\left(\square + m^2\right)\left[\phi\frac{\lambda}{\square}\phi\right] \, \notag\\[4pt]
    &\hspace{12pt} - \frac{1}{6}g \phi^3 -\frac{1}{2}g \phi^2\left(\phi\frac{\lambda}{\square}\phi\right)-\frac{1}{2}g \phi \left(\phi\frac{\lambda}{\square}\phi\right)^2 - \frac{1}{6}g \left(\phi\frac{\lambda}{\square}\phi\right)^3 \notag\\[4pt]
    &\hspace{12pt} + \bar{c} c + \bar{c}\left(\frac{\lambda}{\square} \phi\right) c + \phi\bar{c} \frac{\lambda}{\square}c\,,
\end{align}
where $c$ are the ghosts. 
Let us now show explicitly that correlation functions are unchanged by this field redefinition.
The Feynman vertices originating from redefining the scalar kinetic term are
\vspace{1em}
\begin{subequations}
\label{eq:nonlocal_phi3_feynrules_1}
\begin{align}
\begin{fmffile}{nonlocal_scalar_prop}
\parbox{70pt}{
\begin{fmfgraph}(55,50)
\fmfleft{i1}
\fmfright{o1}
\fmf{plain}{i1,o1}
\end{fmfgraph}}
\end{fmffile} =& \ \frac{i}{p^2-m^2}\,, \\[10pt]\label{eq:rel_ex_arrow}
\begin{fmffile}{nonlocal_phi3}
  \parbox{70pt}{\begin{fmfgraph*}(50,50)
    \fmfleft{i1,i2}
    \fmfright{o1}
    \fmfv{l.a=180,l=$p_2$}{i1}
    \fmfv{l.a=180,l=$p_1$}{i2}
    \fmfv{l.a=0,l=$p_3$}{o1}
    \fmf{plain}{i1,v1}
    \fmf{plain}{i2,v1}
    \fmf{plain}{v1,o1}
    \fmf{phantom_arrow,tension=0,foreground=(.9,,0.2,,0.9)}{v1,o1}
  \end{fmfgraph*}}
\end{fmffile} =& \ i \lambda \bigg[\left(p_3^2-m^2\right)\left( \frac{1}{p_1^2} + \frac{1}{p_2^2} \right)\bigg]\,,\\[20pt] 
\begin{fmffile}{nonlocal_phi4}
  \parbox{70pt}{\begin{fmfgraph*}(55,50)
    \fmfleft{i1,i2}
    \fmfright{o1,o2}
    \fmfv{l.a=180,l=$p_2$}{i1}
    \fmfv{l.a=180,l=$p_1$}{i2}
    \fmfv{l.a=0,l=$p_3$}{o2}
    \fmfv{l.a=0,l=$p_4$}{o1}
    \fmf{plain}{i1,v1}
    \fmf{plain}{i2,v1}
    \fmf{plain}{v1,o1}
    \fmf{plain}{v1,o2}
  \end{fmfgraph*}}
\end{fmffile} =& \ i \lambda^2 \left(\frac{1}{p_1^2}+\frac{1}{p_2^2}\right)\left[(p_1+p_2)^2-m^2\right]\left(\frac{1}{p_3^2} + \frac{1}{p_4^2}\right)\, 
\nn 
&  + (p_2 \leftrightarrow p_3) + (p_2 \leftrightarrow p_4) \, .
\end{align}
\end{subequations}
The arrow on the scalar three-point vertex denotes which leg has the $\Delta^{-1}(p_i) = p_i^2-m^2$ factor in the Feynman rule, as in \cref{sec:perturbative}.
From $\mathcal{L}_{\text{int}}$, we have vertices proportional to $g$:
% \vspace{1em}
{\allowdisplaybreaks
\begin{subequations}
\label{eq:gphi_feynrules}
\begin{align}
\begin{fmffile}{nonlocal_phi3_g_1}
  \parbox{70pt}{\begin{fmfgraph*}(50,50)
    \fmfleft{i1,i2}
    \fmfright{o1}
    \fmfv{decor.shape=circle,decor.filled=30,decor.size=10,foreground=(.1,,0.5,,0.1)}{v1}
    \fmf{plain}{i1,v1}
    \fmf{plain}{i2,v1}
    \fmf{plain}{v1,o1}
  \end{fmfgraph*}}
\end{fmffile} &= i g\,, \label{eq:ex_rel_phi3} \\[1.5em]
\begin{fmffile}{nonlocal_phi3_g_2}
  \parbox{70pt}{\begin{fmfgraph*}(50,50)
    \fmfleft{i1}
    \fmfright{o1,o2,o3}
    \fmfv{decor.shape=circle,decor.filled=30,decor.size=10,foreground=(.1,,0.5,,0.1)}{v1}
    \fmf{plain,tension=3}{i1,v1}
    \fmf{plain}{v1,o1}
    \fmf{plain}{v1,o3}
    \fmf{plain}{v1,o2}
  \end{fmfgraph*}}
\end{fmffile} &= 3i\lambda g \sum_{i=1}^{4}\left(\frac{1}{p_i^2}\right)\,, \\[1.5em]
\begin{fmffile}{nonlocal_phi3_g_3}
  \parbox{70pt}{\begin{fmfgraph*}(50,50)
    \fmfleft{i1,i2}
    \fmfright{o1,o2,o3}
    \fmfv{decor.shape=circle,decor.filled=30,decor.size=10,foreground=(.1,,0.5,,0.1)}{v1}
    \fmf{plain,tension=1.5}{i1,v1}
    \fmf{plain,tension=1.5}{i2,v1}
    \fmf{plain}{v1,o1}
    \fmf{plain}{v1,o3}
    \fmf{plain}{v1,o2}
  \end{fmfgraph*}}
\end{fmffile} &= 6i\lambda^2 g  \sum_{i=1}^{5} \sum_{j\neq i}^{5}\left(\frac{1}{p_i^2}\right) \left(\frac{1}{p_j^2}\right)\,, \\[1.5em]
\begin{fmffile}{nonlocal_phi3_g_4}
  \parbox{70pt}{\begin{fmfgraph*}(50,50)
    \fmfleft{i1,i2,i3}
    \fmfright{o1,o2,o3}
    \fmfv{decor.shape=circle,decor.filled=30,decor.size=10,foreground=(.1,,0.5,,0.1)}{v1}
    \fmf{plain,tension=1}{i1,v1}
    \fmf{plain,tension=1}{i2,v1}
    \fmf{plain,tension=1}{i3,v1}
    \fmf{plain}{v1,o1}
    \fmf{plain}{v1,o3}
    \fmf{plain}{v1,o2}
  \end{fmfgraph*}}
\end{fmffile} &= 6i\lambda^3 g \sum_{i=1}^{6} \sum_{j\neq i}^{6} \sum_{k\neq j \neq i}^{6} \left(\frac{1}{p_i^2}\right) \left(\frac{1}{p_j^2}\right)\left(\frac{1}{p_k^2}\right)\,.
\end{align}
\end{subequations}}
The ghost propagator and interactions from the change in the measure are
\begin{align}
\begin{fmffile}{nonlocal_ghost_prop}
\parbox{70pt}{
\begin{fmfgraph}(55,50)
\fmfleft{i1}
\fmfright{o1}
\fmf{dots_arrow}{i1,o1}
\end{fmfgraph}}
\end{fmffile} =& \ i\,, \\[10pt]
\begin{fmffile}{nonlocal_ccphi}
  \parbox{70pt}{\begin{fmfgraph*}(50,50)
    \fmfleft{i1,i2}
    \fmfright{o1}
    \fmfv{l.a=180,l=$p_2$}{i1}
    \fmfv{l.a=180,l=$p_1$}{i2}
    \fmfv{l.a=0,l=$p_3$}{o1}
    \fmf{dots_arrow}{i1,v1}
    \fmf{plain}{i2,v1}
    \fmf{dots_arrow}{v1,o1}
  \end{fmfgraph*}}
\end{fmffile} =& \ i\lambda \left( \frac{1}{p_1^2}+\frac{1}{p_2^2}\right)\,.
\end{align}
Finally, the coupling to the source becomes
\begin{equation}
\int \dd^4x \, J(x)\phi(x) \rightarrow \int \dd^4 x\, J(x) \left[\phi(x) + \lambda \phi(x) \square^{-1} \phi(x)\right] \, ,
\end{equation}
which gives us two vertices
\begin{subequations}
\label{eq:FR_source_example}
\begin{align}
\begin{fmffile}{source_ex}
\parbox{60pt}{
\begin{fmfgraph*}(40,30)
\fmfcmd{
    path quadrant, q[], otimes;
    quadrant = (0, 0) -- (0.5, 0) & quartercircle & (0, 0.5) -- (0, 0);
    for i=1 upto 4: q[i] = quadrant rotated (45 + 90*i); endfor
    otimes = q[1] & q[2] & q[3] & q[4] -- cycle;
}
\fmfwizard
\fmfleft{i1}
\fmfright{o1}
\fmf{plain,tension=2}{i1,o1}
\fmfv{d.sh=otimes,d.f=empty,d.si=10,label=$J$,l.d=10}{i1}
\end{fmfgraph*}}
\end{fmffile} 
&= 1\,, \\
\begin{fmffile}{source_redef_ex}
\parbox{60pt}{
\begin{fmfgraph*}(40,40)
\fmfcmd{
    path quadrant, q[], otimes;
    quadrant = (0, 0) -- (0.5, 0) & quartercircle & (0, 0.5) -- (0, 0);
    for i=1 upto 4: q[i] = quadrant rotated (45 + 90*i); endfor
    otimes = q[1] & q[2] & q[3] & q[4] -- cycle;
}
\fmfwizard
\fmfright{i2,i3,i4}
\fmfleft{o1}
\fmfblob{0.15w}{o1}
\fmf{plain,tension=1}{i2,o1}
\fmfv{l.a=0,label=$p_1$}{i4}
\fmfv{l.a=0,label=$p_2$}{i2}
\fmf{plain,tension=1}{i4,o1}
\fmfv{d.sh=otimes,d.f=empty,d.si=10,label=$J$,l.d=10}{o1}
\end{fmfgraph*}}
\end{fmffile} &= \lambda \left(\frac{1}{p_1^2}+\frac{1}{p_2^2}\right) \,.
\end{align}
\end{subequations}

Let us now demonstrate how the cancellations proceed in this theory, in the same order as our general proof in \cref{sec:perturbative}.
We will keep all external legs off-shell so that the same cancellations occur within arbitrary loop diagrams.
First, we have
% \vspace{1.5em}
{\allowdisplaybreaks \begin{align}\label{eq:phi3_interaction_feynrules}
% \begin{split}
&\begin{fmffile}{nonlocal_scalar_cancel_1}
\parbox{90pt}{
\begin{fmfgraph*}(70,40)
\fmfleft{i0,i1}
\fmfright{o1,o2}
\fmfv{l.a=180,l=$p_1$}{i1}
\fmfv{l.a=180,l=$p_2$}{i0}
\fmfv{l.a=0,l=$p_4$}{o1}
\fmfv{l.a=0,l=$p_3$}{o2}
\fmf{plain,left=0,tension=2}{v1,v2}
\fmf{phantom_arrow,left=0,tension=0,foreground=(.9,,0.2,,0.9)}{v1,v2}
\fmf{plain,tension=5,label=$q$}{v2,v3}
\fmf{plain,left=0,tension=2}{v4,v3}
\fmf{phantom_arrow,left=0,tension=0,foreground=(.9,,0.2,,0.9)}{v4,v3}
\fmf{plain}{i0,v1}
\fmf{plain}{i1,v1}
\fmf{plain}{v4,o2}
\fmf{plain}{v4,o1}
\end{fmfgraph*}}
\end{fmffile}  + (p_2 \leftrightarrow p_3) + (p_2 \leftrightarrow p_4)\nonumber \\[1.5em]
&= \left[\frac{i( -i \lambda)^2}{q^2-m^2}\right] \bigg[\left(q^2-m^2\right)\left( \frac{1}{p_1^2} + \frac{1}{p_2^2} \right)\bigg] \bigg[\left(q^2-m^2\right)\left( \frac{1}{p_3^2} + \frac{1}{p_4^2} \right)\bigg] + (p_2 \leftrightarrow p_3) + (p_2 \leftrightarrow p_4)\nonumber \\[7pt]
&= -i \lambda^2 \left(\frac{1}{p_1^2}+\frac{1}{p_2^2}\right)\left[q^2-m^2\right]\left(\frac{1}{p_3^2} + \frac{1}{p_4^2}\right) + (p_2 \leftrightarrow p_3) + (p_2 \leftrightarrow p_4) \nonumber\\[1.5em]
&= -\left( \quad \ \ \begin{fmffile}{nonlocal_phi4_tmp}
  \parbox{75pt}{\begin{fmfgraph*}(55,50)
    \fmfleft{i1,i2}
    \fmfright{o1,o2}
    \fmfv{l.a=180,l=$p_2$}{i1}
    \fmfv{l.a=180,l=$p_1$}{i2}
    \fmfv{l.a=0,l=$p_3$}{o2}
    \fmfv{l.a=0,l=$p_4$}{o1}
    \fmf{plain}{i1,v1}
    \fmf{plain}{i2,v1}
    \fmf{plain}{v1,o1}
    \fmf{plain}{v1,o2}
  \end{fmfgraph*}}
\end{fmffile}\right) \, .
% \end{split}
\end{align}}

From $\mathcal{L}_\text{int}$, we have new diagrams proportional to $g$.
The vertex \cref{eq:ex_rel_phi3} gives the physical contribution, and we expect all diagrams involving the new vertices to cancel.
For the diagram proportional to $g \times \lambda$, we have
\vspace{1,5em}
\begin{align}
\begin{split}
\begin{fmffile}{nonlocal_phi3_gphi3}
  \parbox{70pt}{\begin{fmfgraph*}(60,40)
    \fmfleft{i1,i2}
    \fmfright{o1,o2}
    \fmfv{l.a=180,l=$p_2$}{i1}
    \fmfv{l.a=180,l=$p_1$}{i2}
    \fmf{plain,tension=1.5}{i1,v1}
    \fmf{plain,tension=1.5}{i2,v1}
    \fmf{plain,tension=1,label=$q$}{v1,v2}
    \fmf{phantom_arrow,tension=0,foreground=(.9,,0.2,,0.9)}{v1,v2}
    \fmfv{decor.shape=circle,decor.filled=30,decor.size=10,foreground=(.1,,0.5,,0.1)}{v2}
    \fmf{plain,tension=1}{v2,o1}
    \fmf{plain,tension=1}{v2,o2}
  \end{fmfgraph*}}
\end{fmffile} &= (i \lambda) (i g) \sum_{i=1}^{4} \sum_{j > i}^4 \bigg[\left(q^2-m^2\right)\left( \frac{1}{p_i^2} + \frac{1}{p_j^2} \right)\bigg] \left[\frac{i}{q^2-m^2}\right]\\[1.5em]
&= -3i\lambda g \sum_{i=1}^4\left(\frac{1}{p_i^2}\right)\\[1.5em]
&= -\left(\begin{fmffile}{nonlocal_phi3_g_2_tmp}
  \parbox{60pt}{\begin{fmfgraph*}(50,50)
    \fmfleft{i1}
    \fmfright{o1,o2,o3}
    \fmfv{decor.shape=circle,decor.filled=30,decor.size=10,foreground=(.1,,0.5,,0.1)}{v1}
    \fmf{plain,tension=3}{i1,v1}
    \fmf{plain}{v1,o1}
    \fmf{plain}{v1,o3}
    \fmf{plain}{v1,o2}
  \end{fmfgraph*}}
\end{fmffile}\right) \, .
\end{split}
\end{align}
The same pattern follows for diagrams with two and three insertions of the three-point arrow interaction, canceling with the remaining two diagrams in~\cref{eq:gphi_feynrules}.

For diagrams with sources, we have 
\begin{align}
\begin{fmffile}{tree_source_ex}
\parbox{80pt}{
\begin{fmfgraph*}(60,40)
\fmfcmd{
    path quadrant, q[], otimes;
    quadrant = (0, 0) -- (0.5, 0) & quartercircle & (0, 0.5) -- (0, 0);
    for i=1 upto 4: q[i] = quadrant rotated (45 + 90*i); endfor
    otimes = q[1] & q[2] & q[3] & q[4] -- cycle;
}
\fmfwizard
\fmfright{i2,i3,i4}
\fmfleft{o1}
\fmfblob{0.15w}{v1}
\fmf{plain,tension=1}{i2,v1}
\fmf{phantom,tension=1}{i3,v1}
\fmf{plain,tension=1}{i4,v1}
\fmf{plain,tension=2}{v1,o1}
\fmf{phantom_arrow,tension=0,foreground=(.9,,0.2,,0.9),label=$q$}{v1,o1}
\fmfv{l.a=0,label=$p_1$}{i4}
\fmfv{l.a=0,label=$p_2$}{i2}
\fmfv{d.sh=otimes,d.f=empty,d.si=10,l.d=10}{o1}
\end{fmfgraph*}}
\end{fmffile} 
&= i \lambda \left( \frac{1}{p_1^2}+\frac{1}{p_2^2} \right) ( q^2-m^2) \left(\frac{i}{q^2-m^2}\right) \notag\\[3pt]
&= -\lambda \left( \frac{1}{p_1^2}+\frac{1}{p_2^2} \right)\nn[18pt]
&= - \left( \ \begin{fmffile}{tree_source_redef_ex}
\parbox{60pt}{
\begin{fmfgraph*}(40,40)
\fmfcmd{
    path quadrant, q[], otimes;
    quadrant = (0, 0) -- (0.5, 0) & quartercircle & (0, 0.5) -- (0, 0);
    for i=1 upto 4: q[i] = quadrant rotated (45 + 90*i); endfor
    otimes = q[1] & q[2] & q[3] & q[4] -- cycle;
}
\fmfwizard
\fmfright{i2,i3,i4}
\fmfleft{o1}
\fmfblob{0.15w}{o1}
\fmf{plain,tension=1}{i2,o1}
\fmfv{l.a=0,label=$p_1$}{i4}
\fmfv{l.a=0,label=$p_2$}{i2}
\fmf{plain,tension=1}{i4,o1}
\fmfv{d.sh=otimes,d.f=empty,d.si=10,l.d=10}{o1}
\end{fmfgraph*}}
\end{fmffile}\right) \, .
\end{align}
This cancellation is necessary for the invariance of correlation functions, however as discussed in \cref{sec:smatrix}, the situation is simpler for $S$-matrix elements. 
Since our transformation $G[\phi] = \phi \square^{-1} \phi$ does not have the same pole structure as the kinetic term, the vertex \cref{eq:rel_ex_arrow} vanishes as $p_3^2 \rightarrow m^2$ and we may simply ignore the redefinition in the coupling to the source for on-shell external legs.\footnote{This would \textit{not} have been true if we had considered, for example, $G[\phi] = \phi (\square + m^2)^{-1} \phi$.}

Finally, the loops involving ghosts cancel as follows:
\vspace{1.5em}
\begin{align}
\begin{split}
\begin{fmffile}{nonlocal_phi3_tmp}
  \parbox{50pt}{\begin{fmfgraph*}(50,50)
    \fmfleft{i1,i2}
    \fmfright{o1}
    \fmfv{l.a=180,l=$p_2$}{i1}
    \fmfv{l.a=180,l=$p_1$}{i2}
    \fmfv{l.a=90,l=$p$}{o1}
    \fmf{plain}{i1,v1}
    \fmf{plain}{i2,v1}
    \fmf{plain}{v1,o1}
    \fmf{phantom_arrow,tension=0,foreground=(.9,,0.2,,0.9)}{v1,o1}
  \end{fmfgraph*}}
\end{fmffile} \ \
\times \ \
\begin{fmffile}{nonlocal_tmp2}
\parbox{70pt}{
\begin{fmfgraph}(60,40)
\fmfleft{i1}
\fmfright{o1}
\fmf{plain}{i1,o1}
\end{fmfgraph}}
\end{fmffile} &= i\lambda \left[(p^2-m^2)\left(\frac{1}{p_1^2}+\frac{1}{p_2^2}\right)\right] \left[\frac{i}{p^2-m^2}\right] \\ 
&= -\lambda \left(\frac{1}{p^2_1}+\frac{1}{p_2^2}\right)\\[1.5em]
&= 
\begin{fmffile}{nonlocal_phi3_tmp_gh}
  \parbox{50pt}{\begin{fmfgraph*}(50,50)
    \fmfleft{i1,i2}
    \fmfright{o1}
    \fmf{dots_arrow}{i1,v1}
    \fmf{plain}{i2,v1}
    \fmf{dots_arrow}{v1,o1}
  \end{fmfgraph*}}
\end{fmffile} \ \ 
\times \ \
\begin{fmffile}{nonlocal_tmp_gh_2}
\parbox{70pt}{
\begin{fmfgraph}(60,40)
\fmfleft{i1}
\fmfright{o1}
\fmf{dots_arrow}{i1,o1}
\end{fmfgraph}}\,,
\end{fmffile}
\end{split}
\end{align}
and so the $(-1)$ from any closed ghost loop causes it to cancel the corresponding scalar loop.
We should emphasize that ghost loops do not vanish for this redefinition and so they must be included.
For example, the one-loop correction to the two-point function includes the diagram
\begin{align}
\begin{fmffile}{twopoint_ghost}
\parbox{80pt}{
\begin{fmfgraph*}(80,60)
       \fmfleft{i}
       \fmfright{o}
       \fmfv{label=$p$}{i}
       \fmf{plain,tension=1}{i,v1}
       \fmf{plain,tension=1}{v2,o}
       \fmf{dots_arrow,left,tension=0.4,label=$p-k$}{v2,v1}
       \fmf{dots_arrow,left,tension=0.4,label=$k$}{v1,v2}
    \end{fmfgraph*}}
\end{fmffile} &= -\lambda^2\int \frac{\dd^4k}{(2\pi)^4} \left(\frac{1}{p^2}+\frac{1}{k^2}\right)\left(\frac{1}{p^2}+\frac{1}{(p-k)^2}\right)\,,
\end{align}
which is nonzero even when using dimensional regularization.

This accounts for all of the new terms in \cref{eq:rel_ex_Lagr} generated by the redefinition, showing invariance of correlation functions and $S$-matrix elements.

\subsection{A Useful Nonlocal Field Redefinition}

Let us now turn to a nonrelativistic example.  This closely follows the discussion in Ref.~\cite{Brauner:2024juy}. 
We start with the Lagrangian for a single complex Schr\"{o}dinger field:
\begin{align}
    \Lagr = 2iM \phi^{*} \partial_{0} \phi - \nabla \phi^{*} \cdot \nabla \phi + m^2 \phi^{*}\phi - \lambda (\phi^{*}\phi)^{2} \,.
\end{align}
We expand around the vev:
\begin{align}
\phi = \frac{e^{i (\theta + \pi(x)/v)}}{\sqrt{2}} \big( v + \chi(x)\big )\,,
\end{align} 
where $v=m/\sqrt{\lambda}$ and $\theta$ is an arbitrary phase. The Lagrangian now takes the form
\begin{align}
    \Lagr =& - 2M \chi \partial_{0} \pi - \frac{1}{2} (\nabla \chi)^2 - m^2 \chi^2 - \frac{1}{2} (\nabla \pi)^2 \nn[4pt]
    & - \lambda v \chi^{3} - \frac{\lambda}{4} \chi^{4} - \frac{M}{v} \chi^{2} \partial_{0} \pi - \left( \frac{\chi}{v} + \frac{\chi^{2}}{2v^2} \right) (\nabla \pi)^{2} \,,
\end{align}
up to constant terms and total derivatives. The quadratic part of the Lagrangian mixes the two fields $\pi$ and $\chi$. No local field redefinition can diagonalize the quadratic part of the Lagrangian. Nevertheless, we can extract the dispersion relation by Fourier transforming the bilinear part of the Lagrangian and setting the determinant of the resulting matrix to zero. This gives
\begin{align}
    \label{eq:dispersion_pi}
    E(p) = \frac{1}{2M}\sqrt{\boldsymbol{p}^2 (\boldsymbol{p}^2 + 2m)} \,,
\end{align}
where $E$ is the energy and $\boldsymbol{p}$ is the three-momentum.

Equipped with nonlocal field redefinitions, we can obtain this result directly at the level of the Lagrangian. Consider the following field redefinition:
\begin{align}
    \chi \rightarrow \tilde{\chi} =  \chi + \frac{1}{\nabla^2 - 2m^2} 2M \partial_{0} \pi ,
\end{align}
which diagonalizes the quadratic part of the Lagrangian,
\begin{align}
    \Lagr_{2} = \frac{1}{2} \chi (\nabla^2 - 2m^2) \chi - \frac{1}{2} (2M \partial_{0}\pi) \frac{1}{\nabla^2 - 2m^2} (2M \partial_{0}\pi) - \frac{1}{2} (\nabla \pi)^2 \,.
\end{align}
Now we can directly read off the dispersion relation for $\pi$, which is in agreement with \cref{eq:dispersion_pi}.

\subsection{A Useful Time-dependent Field Redefinition}
In Ref.~\cite{Mojahed:2022nrn}, the following Lagrangian for the field $\theta$ was considered:
\begin{align}
    \Lagr = \frac{1}{8\tilde c_4} \left[ 1 - 2c_3\partial_{0}\theta - \sqrt{(1-2c_3\partial_{0}\theta)^2 - 8 \tilde c_4 \big[ (\partial_{0}\theta)^2 - (\nabla\theta)^2 \big]} \right]\,.
\end{align}
The Lagrangian is controlled by two parameters, $c_3$ and $\tilde c_4$. This theory appears to be nonrelativistic. However, all scattering amplitudes are independent of the coupling $c_3$ and equal to the scattering amplitudes of the relativistic Dirac-Born-Infeld (DBI) theory \cite{Mojahed:2022nrn}. We can understand this fact by changing field basis:
\begin{align}
    \label{eq:fluidFR}
    \theta \rightarrow \tilde{\theta} = \alpha\theta + \beta t\,, 
\end{align}
where 
\begin{align}
%\alpha^2 = \left(\frac{2\tilde c_4}{2 \tilde c_4-c_3^2 }\right)^{3/2}\,, 
\alpha = \left(\frac{2\tilde c_4}{2 \tilde c_4-c_3^2 }\right)^{3/4}\,, 
    \qquad \text{and}\qquad
    \beta = \frac{c_3}{2(c_3^2 - 2 \tilde c_4)} \,.
\end{align}
After dropping constant terms and total derivatives, the Lagrangian becomes
\begin{align}
    \Lagr = - \frac{1}{\kappa}\sqrt{1- \kappa \big[ (\partial_0 \theta)^2 - c^2 (\nabla\theta)^2\big] }  \,,
\end{align}
where $\kappa = 4\sqrt{2\tilde c_4(2\tilde c_4 - c_3^2)}$ and $c^2 = \tfrac{2\tilde c_4}{2\tilde c_4 - c_3^2}$. This is the DBI theory, which also coincides with the theory of an exceptional fluid with an enhanced soft limit \cite{Cheung:2023qwn}. This was noted in Ref.~\cite{Mojahed:2022nrn}, but the legality of the field redefinition in \cref{eq:fluidFR} was questioned since it depends on time explicitly. However, the diagrammatic proof of field-redefinition invariance presented in \cref{sec:proof} ensures that this is a valid field redefinition.

\section{Outlook}\label{sec:Outlook}

In this paper, we have explored the space of allowed field redefinitions for EFTs.
We have shown how field-redefinition invariance is manifest at the level of correlation functions and $S$-matrix elements after LSZ reduction.
For linear and field-independent transformations, we have additionally demonstrated how field-redefinition invariance remains even after resumming modifications into the propagator.
Our focus was on unconventional field redefinitions that are typically not considered.
The space of allowed redefinitions is vast: symmetry breaking, nonlocality, and in certain cases explicit coordinate dependence all present no obstacles. 

There are many future directions to explore.  One immediate application is to understand the implications for \emph{field-space geometry}, where derivative-independent field redefinitions are identified as coordinate changes on a field manifold.  This subject has a long history, beginning with a geometric formulation of nonlinear sigma models~\cite{Honerkamp:1971sh, Volkov:1973vd, Tataru:1975ys, Alvarez-Gaume:1981exa, Alvarez-Gaume:1981exv, Vilkovisky:1984st, DeWitt:1984sjp, Gaillard:1985uh, DeWitt:1985sg}. Recently, this idea has seen a renaissance in the context of more general EFTs~\cite{Alonso:2015fsp, Alonso:2016btr, Alonso:2016oah, Nagai:2019tgi, Helset:2020yio, Cohen:2020xca, Cohen:2021ucp, Alonso:2021rac, Banta:2021dek, Talbert:2022unj, Alonso:2023jsi,Finn:2019aip, Finn:2020nvn, Cheung:2021yog, Alonso:2022ffe, Helset:2022tlf, Helset:2022pde, Assi:2023zid, Jenkins:2023rtg, Jenkins:2023bls, Gattus:2023gep, Alonso:2023upf, Gattus:2024ird,Derda:2024jvo,Helset:2024vle,Li:2024ciy}. As we have emphasized in this work, the space of valid field redefinition is much larger than those that are captured by the standard field-space-geometry picture. The geometric EFT framework must correspondingly be generalized. As of now, several attempts have been made to accommodate (local) field redefinitions with derivatives~\cite{Cohen:2022uuw, Cheung:2022vnd, Craig:2023wni, Craig:2023hhp, Alminawi:2023qtf, Cohen:2024bml, Lee:2024xqa}. These ideas are still in their infancy, and now we see that they must be generalized even further to account for nonlocal field redefinitions. 
Also, from the discussion of tadpole resummations in \cref{sec:Resummation}, we can contemplate novel ways to compute $S$-matrix elements at various points in field space.
We are optimistic that this paper will inspire further progress in these directions.

This paper emphasizes the importance of the often neglected coupling between the field and the external source. The conventional linear coupling is specific to a particular choice of field basis, and one generically induces nonlinear source couplings through field redefinitions. This naturally leads one to wonder if there exists an optimal choice of coupling involving the source. One famous example of exploiting this coupling is the Vilkovisky-deWitt effective action, which is obtained after modifying the source term to keep the field-space covariance manifest for off-shell computations~\cite{Vilkovisky:1984st,Batalin:1987fx}. Similarly, renormalization-group and matching computations are geometrized with a judicious choice of source term \cite{Alonso:2016oah,Helset:2022pde,Assi:2023zid,Li:2024ciy}. We expect that other modifications of the source term will be of use in future work. 

In practice, it may not always be obvious how to choose the appropriate coupling to the source.
For example, it was demonstrated in the context of SCET that there is a preferred choice to reproduce the expected off-shell QCD amplitudes for renormalization~\cite{Beneke:2019kgv}.
It would be interesting to further explore different choices for the coupling to the source in SCET and other modern EFTs, as well as the interplay between field redefinitions and renormalization in general.

Modifying the source term may also hold the key for defining observables free of infrared singularities. The Faddeev-Kulish dressing \cite{Kulish:1970ut} may turn out to be equivalent to an appropriately-chosen source term. In the same spirit, recent progress towards understanding the general structure of infrared singularities for the $S$-matrix~\cite{Frye:2018xjj,Hannesdottir:2019umk,Hannesdottir:2019opa} may be informed by the results in this work.

Nonlocal field redefinitions are potentially very useful in EFTs with a preferred frame, such as EFTs for nonrelativistic systems~\cite{Leutwyler:1993gf,Leutwyler:1996er,Son:2002zn,Andersen:2002nd,Brauner:2010wm,Nicolis:2011pv,Nicolis:2012vf,Nicolis:2013lma,Nicolis:2015sra,Cuomo:2020gyl,Nicolis:2022llw,Creminelli:2022onn,Hui:2023pxc,Creminelli:2023kze,Creminelli:2024lhd,Green:2022slj,Cheung:2023qwn,Grall:2020ibl}, the EFT of inflation~\cite{Creminelli:2006xe,Cheung:2007st}, and EFTs for black-hole binaries \cite{Kosmopoulos:2023bwc,Cheung:2023lnj}.
We already saw a couple of examples in \cref{sec:Examples}, and we anticipate a host of other applications.  
With nonlocal field redefinitions at our disposal, it might be possible to dramatically simplify the analysis of these interesting theories.

Field redefinitions are an essential tool for the study of EFTs, both conceptually and with a wide range of practical applications.  
Now that we have extended the space of allowed field redefinitions, we can only begin to anticipate the multitude of future implications.  
Looking forward, we are very excited to see what the community will discover by playing games with field redefinitions without the prejudice of locality and symmetry.

\subsection*{Acknowledgments}

We thank Joydeep Chakrabortty, Xiaochuan Lu, Matthew McCullough, George Sterman, Robert Szafron, Zhengkang Zheng, and Sasha Zhiboedov for useful discussions.
The work of T.~C.~is supported by the U.S.~Department of Energy grant DE-SC0011640.
The work of M.~F.~is supported by the U.S.~National Science Foundation grant PHY-2210533. 
M.F. is also grateful for the support of the U.S. Department of Energy, Office of Science, Office of Workforce Development for Teachers and Scientists, Office of Science Graduate Student Research (SCGSR) program, during which this work was initiated while visiting CERN.
The SCGSR program is administered by the Oak Ridge Institute for Science and Education (ORISE) for the DOE. ORISE is managed by ORAU under contract number DE-SC0014664.

\end{spacing}

\begin{spacing}{1.09}
\addcontentsline{toc}{section}{\protect\numberline{}References}%
\bibliographystyle{JHEP}
\bibliography{fieldRedefinitions}
\end{spacing}

\end{document}